\documentclass[11pt]{article}
\usepackage{eurosym,float}
\usepackage[utf8]{inputenc}
\usepackage[T1]{fontenc}
\usepackage{fourier}   
\usepackage[a4paper,top=3cm,bottom=3cm]{geometry}
\usepackage{amsmath,amssymb}
\usepackage{amsthm}
\usepackage{dcolumn}
\usepackage{booktabs, float, caption, makecell}
\usepackage{siunitx}
\usepackage[flushleft]{threeparttable}
\usepackage{lscape}
\usepackage{graphicx}
\usepackage{multirow}
\usepackage{setspace}
\usepackage{hyperref}
\usepackage{pgfplots}
\usepackage{authblk}
\usepackage{bm}
\usepackage[numbers,square,sort]{natbib}
\usepackage{textalpha}
\usepackage{diagbox}
\usepackage{ragged2e}
\justifying
\usepackage{soul}
\usepackage{textcomp}
 
\usepackage{xfrac}
\usepackage{caption}
\usepackage{subcaption}
\usepackage{xcolor}
\usepackage{comment}
\usepackage{array}
\usepackage{chngcntr}
\usepackage[normalem]{ulem}
\setcitestyle{authoryear,open={(},close={)}}
\pgfplotsset{compat=1.18}

\newcommand{\angelos}[1]{\textcolor{blue}{Angelos: #1}}
\newcommand{\ilias}[1]{\textcolor{red}{Ilias: #1}}

\newcolumntype{d}[1]{D{.}{.}{#1}}

\begin{document}
	
\title{The heterogeneous causal effects of the EU's Cohesion Fund\thanks{We would like to thank Jan Ditzen, George Kapetanios, Yiannis Karavias, Nicholas Lazarou, Simone Salotti, Pietro Santoleri, John Tsoukalas, participants of the TEDAM seminar series at the 21st Conference on Research on Economic Theory and Econometrics, the 64th European Regional Science Association Congress and the 9th Applied Macro and Empirical Finance Conference. The views expressed in this paper are those of the authors alone and do not represent the official views of the European Commission. All remaining errors are our own.}}

\author[1]{Angelos Alexopoulos 
\thanks{\textbf{Corresponding author:} \href{mailto:angelos@aueb.gr}{angelos@aueb.gr}}}
\author[2]{Ilias Kostarakos\thanks{\href{mailto:ilias.kostarakos@ec.europa.eu}{ilias.kostarakos@ec.europa.eu}}}
\author[1]{Christos Mylonakis\thanks{\href{mailto:mylonakisc@gmail.com}{mylonakisc@gmail.com}}}
\author[1]{Petros Varthalitis\thanks{\href{mailto:pvarthalitis@aueb.gr}{pvarthalitis@aueb.gr}}}

\affil[1]{Athens University of Economics and Business, Department of Economics, Athens, Greece}
\affil[2]{European Commission, Joint Research Centre}
\date{\today}
	
\maketitle
	
\begin{abstract}
This paper quantifies the causal effect of cohesion policy on regional output and investment across the EU regions, focusing on one of its least studied instruments, namely, the Cohesion Fund (CF).  The novelty of the paper lies in deviating from the standard methodologies employed in the existing literature such as regression discontinuity design (RDD) or instrumental variables (IV). Instead, we adopt a recently developed causal inference approach based on matrix completion, implemented within a factor model framework. This allows us to propose a novel framework for evaluating the effectiveness of the CF as a key EU cohesion policy instrument. Specifically, we estimate the time varying distribution of the CF’s causal effects across EU regions and derive key distribution metrics useful for policy evaluation. Our analysis shows that relying solely on average treatment effects masks significant heterogeneity and can lead to misleading conclusions about the effectiveness of the EU's cohesion policy. We find that the impact of the CF is frontloaded, peaking within the first seven years following a region’s initial inclusion in the program. The distribution of the effects during this first seven-year cycle of funding is right-skewed with relatively thick tails. This indicates generally positive effects but unevenly distributed across regions. Moreover, the magnitude of the CF’s effect is inversely related to a region’s relative position in the initial regional GVA distribution, i.e., relatively poorer recipient regions experience higher effects compared to relatively richer regions. We find a non-linear relationship with diminishing returns, whereby the impact of the CF begins to decline beyond a certain threshold as the ratio of CF funds received to a region’s gross value added (GVA) increases.
\end{abstract}
\vspace{1cm}

\newpage
\section{Introduction}\label{sec:intro}
Cohesion policy is one of the most important fiscal policy initiatives at the European Union (EU) level. Over the last three decades, nearly \euro 960 billion (in 2015 prices) have been disbursed to the EU regions as part of this sub-national transfers program. To put things in a historical perspective, cohesion-related expenditures account for almost one-third of the EU's budget (see \cite{bachtler} and \cite{benedetto}), making it one of the largest redistribution programs ever implemented in the EU. In the current seven-year budget period,\footnote{The so-called Multiannual Financial Framework - see the Council Regulation 2020/2093 for additional details on the expenditure breakdown for the 2021-2027 budget.} \euro392 billion have been dedicated to cohesion policy objectives, reinforcing its status as a top spending priority of the EU. Naturally, given its scale and overarching objectives, the effectiveness of cohesion policy has been debated over time by both academics and policymakers (see \citet{ec2024}).

This paper quantifies the causal effect of EU cohesion policy on regional output (Gross Value Added) and investment (Gross Fixed Capital Formation). We focus on one of the least studied instruments of cohesion policy: the Cohesion Fund (CF). This policy instrument is available only to the relatively poorer Member States and has accounted for nearly 20\% of the total cohesion policy expenditures over time. The first novelty of the paper is the adoption of recent developments in causal inference, i.e., \cite{xu2017generalized}, \cite{athey2021matrix} and \cite{borusyak2024revisiting} to study this type of questions.\footnote{For a recent review see \cite{xu2023causal}} Thus, we extend beyond the existing literature, see for example \cite{becker1}, \cite{becker3}, \citet{gagliardi} and \citet{crescenzi} which has primarily focused on estimating a common average treatment effect across all treated regions. In contrast, our empirical strategy enables us to capture not only the average treatment effect of cohesion policy, but also its heterogeneous effects across regions and over time. To this end, we estimate the entire time-varying distribution of regional treatment effects. 

This perspective is particularly relevant for evaluating the core objectives of the CF, which are not only to increase the growth of average regional income but, primarily, to reduce economic disparities within the EU. Consequently, the second novelty of this paper is the introduction of standard distributional statistics as a central tool to evaluate the performance of the CF against its policy objectives. Specifically, we employ measures such as the share of regions with positive treatment effect, skewness, kurtosis, average treatment effect by quintile, and the regional Gini coefficient of the distribution over time (e.g., successive seven-year funding cycles).

Building on this framework, we can evaluate the performance of CF in several complementary dimensions with respect to the existing literature (see e.g., \citet{becker1}, \citet{becker3} and \citet{crescenzi}). To make our results directly comparable to this literature, we start with the average impact on key economic outcomes, such as output and investment. We then analyze the evolution of the distribution of effects to infer convergence dynamics and shifts in regional disparities to identify which types of region benefit most. Finally, we investigate how policy effectiveness varies with transfer intensity, shedding light on potential non-linearities and offering guidance for the optimal design and redistribution of future cohesion interventions.

The empirical literature evaluating the causal effects of EU cohesion policy has so far relied predominantly on Regression Discontinuity Designs or instrumental variable IV strategies. These approaches exploit institutional eligibility thresholds — such as the $75\%$ GDP per capita rule defining access to Objective$~1$ or the Convergence Objective or `less-developed' status\footnote{Objective 1 or the Convergence Objective or `less-developed' status refers to the classification scheme of the European Commission that determines which regions will receive the lion's share of the available structural funds. As elaborated in the Council Regulation 2081/93 and the European Commission Decisions 1999/502/EC, 2006/595/EC and 2014/99/EC regions below the threshold of 75\% of the average EU GDP per capita belong to this class.} — to identify the local causal effects of cohesion transfers on outcomes such as output, investment, or employment (see, e.g., \citealt{becker1,becker3,becker4}, \citet{pellegrini} and \citet{cerqua}). While such a design provides credible identification, it is typically limited to local average treatment effects for regions close to the eligibility threshold and offers little insight into the dynamic, heterogeneous, or distributional consequences of cohesion spending.

Our focus shifts towards an event-study-type framework, which is more suitable for evaluating dynamic treatment effects and for capturing heterogeneity along several dimensions, e.g., time, regional income levels, transfer intensity and space. However, applying such an approach to the evaluation of cohesion policy raises several empirical challenges. In particular, regional data are characterized by strong spatial and cross-sectional dependencies, while policy impacts are likely to be heterogeneous across regions with different structural characteristics. These complexities must be addressed to obtain credible causal estimates and fully understand the policy’s distributional consequences.

A first empirical challenge arises from spatial dependence across regions, typically driven by geographical proximity and local spillovers. Ignoring these interdependencies can bias treatment effect estimates, as neighboring regions often exhibit correlated outcomes. Several studies have employed standard spatial econometric techniques to address this issue (see, among others, \citet{mohl},  \citet{bourdin}, \citet{crescenzi}, \citet{dicaro} and \citet{amendolagine}), although applications in a causal inference context remain rare, with \citet{crescenzi} being a notable exception through their spatial RDD approach. A second challenge is cross-sectional dependence, which reflects broader, system-wide sources of correlation across regions. This dependence is often due to unobserved common factors, such as the 2008 Global Financial Crisis or the European sovereign debt crisis, whose effects extend well beyond local neighbourhoods. When these factors are correlated with treatment assignment, they can lead to substantial estimation biases if not properly addressed. This dependence is often due to unobserved common factors, such as the 2008 Global Financial Crisis or the European sovereign debt crisis, whose effects extend well beyond local neighbourhoods. Importantly, these shocks can bias conventional estimators not only when they are correlated with treatment assignment, but also when regions exhibit differential exposure to the same aggregate shock, leading to heterogeneous comovements in outcomes. Both channels generate substantial estimation bias if not properly accommodated.

Finally, regional policies are unlikely to exert homogeneous effects across space. Yet, much of the existing literature implicitly assumes identical treatment effects across regions, overlooking important heterogeneity linked to income levels, structural characteristics, and initial economic conditions. Recognising and modelling such heterogeneity is crucial not only for understanding the full distributional consequences of cohesion policy but also for assessing how its effectiveness evolves over time and across different types of regions. Our approach explicitly addresses these issues by estimating region-specific, time-varying effects, thereby providing new insights into the differential impacts of cohesion spending and informing the design of more targeted and effective policy interventions.

We address these challenges by employing a matrix completion approach implemented within a factor model framework \citep{xu2017generalized, athey2021matrix}. This approach generalizes both the conventional Two-Way Fixed Effects (TWFE) model and the synthetic control (SC) method \citep{abadie2010synthetic}, thereby combining their main advantages while relaxing key identifying assumptions. In particular, by modelling the data-generating process through interactive fixed effects \citep{pesaran, bai}, we can explicitly account for unobserved time-varying common factors — potentially correlated with treatment assignment — that would otherwise bias conventional DiD estimates \citep{gobillon2016regional}. This feature is especially relevant in our setting, where regional data are relatively scarce and where such confounders, if ignored, would invalidate the parallel trends assumption central to standard DiD approaches \citep{angrist2009mostly, sun2021estimating}, \citet{de2024difference}. Our framework also extends beyond classical synthetic control methods by allowing for staggered treatment adoption across multiple regions and for heterogeneous factor loadings, which capture region-specific responses to common shocks. As a result, we are able to recover region-specific, time-varying treatment effects in the presence of complex cross-sectional dependencies and heterogeneous dynamics — an innovation in the context of cohesion policy evaluation. This richer modelling approach not only strengthens causal identification but also enables a more nuanced assessment of the Cohesion Fund’s effectiveness across time, space, and the income distribution.

Our findings show that, on average, the CF has a sustained positive impact on both GVA and GFCF in recipient regions, relative to the counterfactual where the EU does not provide funds. This is consistent with previous research, see for example, \citet{becker1} and \citet{pellegrini} who document a positive impact of the structural funds on growth. The CF effect is front-loaded, reaching its peak within the first seven years after a region is included in the program. For the average `hypothetical' treated EU region, GVA per capita is about \euro815 higher after seven years and \euro2961 higher after fourteen years than in the counterfactual scenario. GFCF per capita is \euro395 and \euro680 higher over the same periods, respectively.

These findings are further supported by distributional metrics such as the average treatment effect by quintile and what we call the regional Gini coefficient. Quintiles are defined by regional GVA per capita in the year prior to treatment.  We estimate that regions in the lowest (highest) quintile exhibit the largest (smallest) average treatment effects. Similarly, the regional Gini coefficient—which captures cross-regional dispersion in GVA per capita—declines from about 31\% to 28\% over the period 1995–2022. In sharp contrast, under the counterfactual where we assume the absence of CF program, there is no reduction in the dispersion of GVA across regions; in fact, we estimate a modest increase over time. Similar patterns emerge for investment (GFCF), with even stronger and more uneven effects that are highly concentrated among less developed regions.

Finally, we find that the relationship between treatment intensity, measured as the share of CF expenditures to GVA in real terms, and economic impact is clearly non-linear across all time horizons.  After the seventh year under treatment, the optimal CF intensity is approximately 0.6\% of GVA, at which point the estimated effect reaches 25\%. Beyond this threshold, additional funding produces diminishing economic returns. When the horizon is fourteen or twenty one years the optimal intensity is estimated at 0.7\% and 0.86\%, yielding maximum effects of 0.32\% and 0.47\%, respectively. These results support the hypothesis of a maximum desirable level of transfers proposed by \cite{becker2}, suggesting that while moderate levels of EU funding can stimulate growth, when transfers exceed a threshold, they can become inefficient or even counterproductive. 

The rest of our paper is laid out as follows: section~\ref{sec:literature} presents an overview of the related literature while section~\ref{sec:background} presents some additional information on the Cohesion Fund. Section~\ref{sec:data} presents our dataset, while section~\ref{sec:empirical} analyses our methodology and presents the empirical model. Section~\ref{sec:results} presents the results and section~\ref{sec:conclusions} concludes.
	
\section{Related literature}\label{sec:literature} 
The present paper contributes to the empirical literature on the ex-post evaluation of cohesion policy transfers. In recent years, this literature has focused on the application of causal inference techniques.\footnote{There exists a large empirical literature that has employed more traditional (spatial) econometric approaches to identify the effects of cohesion policy, focusing either on the long-run effects or on the calculation of multipliers. More details on this strand are available in Appendix~\ref{app:litreview} and the bottom panel of Table~\ref{tab:litreview}} By calculating counterfactual scenarios and comparing them with the actual observed outcomes, they arrive at the net causal impact of the cohesion transfers. A schematic review of the relevant literature can be found in the top panel of Table~\ref{tab:litreview}.

Following the seminal contribution of \citealt{becker1}, a number of papers have leveraged on a specific institutional characteristic of the cohesion transfers programme to setup their research design. Specifically, the European Commission classifies regions as attaining the `Objective I' or `Convergence Objective' or `less developed' status if their GDP per capita (in PPS) prior to the commencement of each budget cycle is less than 75\% of the EU average.\footnote{Indicatively, for the 2014-2020 period the list of regions belonging to the `less developed' group was decided using data for the 2007 - 2009 period, as detailed in the Commission Implementing Decision 2014/99.} This classification in turn determines the amount of cohesion transfers each region is eligible for;
the `less developed' regions receive the bulk of the transfers.\footnote{As such, all regions receive at least some funding; eligibility or not for the `less developed' status determines the size of the transfer.}  As such, the focus is on whether being assigned the `less developed status has a causal impact on economic performance. 

The analysis of \citealt{becker1, becker2, becker4}, \citealt{gagliardi} and \citealt{crescenzi} rests on a fuzzy\footnote{The fuzzy design arises due to the fact that some regions that were formally eligible for receiving the `less developed' status were not included in this group of regions and vice versa.} RDD setup and, broadly, points to an overall positive causal effect. A positive effect is also identified in \citet{pellegrini} who use a `sharp' RDD approach. Interestingly, all these contributions examine the impact of all the available policy instruments simultaneously\footnote{That is, the European Regional Development Fund, the European Social Fund and the Cohesion Fund.} and largely do not distinguish between programming periods. Lastly, in an interesting extension, \citet{becker3} estimate a heterogeneous local treatment effect based on an extension of the standard RDD that allows for the examination of the impacts of each region's absorptive capacity.

A separate branch of this literature shifted the focus to the more granular issue of whether the level of transfer intensity that regions receive is what matters for economic performance; that is, whether higher levels of treatment lead to a higher or lower impact on the outcome compared to lower levels of treatment. In this context, \citet{hagen}, \citet{becker2} and  \citet{cerqua} employ a generalized propensity score approach following \citet{hirano2004} and estimate the so-called \textit{dose-response} function. This allows the examination of whether there exists either a \textit{minimum necessary} level or a \textit{maximum desirable} level of regional transfer intensity necessary to achieve positive effects from cohesion policy expenditures. Both \citet{becker2} and  \citet{cerqua} identify the presence of non-linear inverted-U shaped relationship between transfers and economic performance. 

Our work complements and extends these contributions by adopting a generalized synthetic control framework rooted in the matrix completion and interactive fixed effects literature (\citealt{gobillon2016regional}, \citealt{athey2021matrix}; \citealt{xu2017generalized,xu2023causal}). This approach addresses several well-known limitations of RDD-, IV-, and dose–response-based evaluations. First, whereas RDD and IV designs identify only local or complier-specific effects around the eligibility threshold, our estimator yields \emph{region- and time-specific} causal effects for the full set of treated units. This enables the analysis of the dynamic evolution of policy impacts over the entire treatment horizon and across the full distribution of regions, rather than a single average homogeneous effect. Second, unlike dose–response models, our estimator does not require unconfoundedness or common support conditions and avoids the high finite-sample variance associated with nonparametric identification. Third, by incorporating interactive fixed effects, our framework explicitly models unobserved time-varying confounders and heterogeneous exposure to common shocks, such as the 2008 Global Financial Crisis or the European sovereign debt crisis; these are factors that can bias conventional RDD/IV or DiD estimators but these approaches cannot adequately accommodate.

Beyond these methodological advantages, our framework enables us to exploit the full 1993–2022 period during which the Cohesion Fund has operated and to obtain \emph{annual} causal effects rather than effects aggregated over programming periods. This allows us to examine the dynamics of adjustment, assess the extent to which Cohesion Fund transfers mitigated large macroeconomic shocks, and study whether the relationship between transfer intensity and regional outcomes exhibits a non-linear pattern. Finally, by focusing on the Cohesion Fund---the instrument targeted at the EU's relatively poorer Member States—we shed new light on the heterogeneous effectiveness of cohesion spending across space, time, and the regional income distribution.

We note here that there exists a separate, voluminous empirical literature related to the identification of the impact of cohesion policy. This literature focuses on more traditional (spatial) econometric approaches and is surveyed in Appendix~\ref{app:litreview}; for a schematic review see the bottom panel of Table~\ref{tab:litreview}. Indicatively, \citet{canova} provide evidence on the size of the multipliers of the cohesion policy instruments, noting the heterogeneity arising due to, for example, location and the level development; \citet{amendolagine} employ a heterogeneous panel model and show that the impact of cohesion Policy is positive but highly heterogeneous, while also generating spatial spillover effects.

\section{Background: the Cohesion Fund}\label{sec:background}

The European Structural and Investment Funds (henceforth, ESIF) constitute a major component of a system of (sub)national transfers made by the EU to the Member States to foster regional economic performance. ESIF is primarily composed by the European Regional Development Fund (ERDF), the European Social Fund (ESF) and the Cohesion Fund (CF) --which are collectively known as the cohesion policy instruments-- and the European Agricultural Fund for Rural Development (EAFRD) (which, together with the European Agricultural Guarantee Fund (EAGF) are the two components of the Common Agricultural Policy (CAP)). 

The vast majority of the relevant literature has focused either on the impact of all cohesion policy instruments combined or on the two largest instruments, namely the ERDF and the ESF. In this paper we deviate from this trend and focus our attention on the largely neglected CF instrument.

The CF was established in 1994 in order to further strengthen the cohesion of the EU via funding investment projects related to the fields of environment and transport infrastructure. A key characteristic of the Cohesion Fund is that the rule based on which the European Commission allocates the budget across the Member States differs compared to the rest of the ESIF instruments. In particular, the fund is available to countries that exhibit per capita Gross National Income (GNI) lower than 90\% of the EU27 average. As a result, the CF instrument is targeting, by design, the poorest European territories and offers them additional assistance to achieve higher levels of output. This is made evident in the following Table~\ref{tab:cf_descriptives} which breaks down CF expenditures by programming period and also provides further details about the fund's recipients.

\begin{table}[H]
\centering
\caption{Cohesion Fund recipients and transfers}
\label{tab:cf_descriptives}
\begin{threeparttable}
\begin{tabular}{ccccccc}
\toprule
\centering
Period& Countries& Regions& GDP pc <75\%& Population (\%)& CF amount& Share of CP(\%) \\
\midrule
1994-1999 & 4 & 42& 16& 14.9& 18078.27& 25.04 \\
2000-2006 & 16& 100& 74& 37.3& 30618.79& 17.7\\
2007-2013 & 16& 102& 71& 39.6& 68939.32& 27.1\\
2014-2020 & 15& 83& 79& 38.9&  47953.53& 22.3\\
\bottomrule
\end{tabular} 
\begin{tablenotes}
\scriptsize
\item \textit{Notes}: Values are in \euro millions. 
\item \textit{Notes}: The dates in the table depict the beginning of the respective programming period.
\item The number of regions with GDP per capita lower than 75\% of the EU average is calculated based on data corresponding to the first year of each programming period. 
\item Share of CP refers to CF payments as a share of the total cohesion policy payments over the respective period. 
\end{tablenotes}
\end{threeparttable}
\end{table}

As can be observed in the Table, the scope of the CF increased significantly over time. In particular, when the Fund first became operational, payments were made toward four countries only --namely, Greece, Ireland, Portugal and Spain-- that accounted for 15\% of the EU's population. Moreover, only 40\% of the beneficiaries exhibited a level of GDP per capita lower than 75\% of the EU average, effectively classifying them as being `less developed' regions. 
During the 2000-2006 period, which saw the enlargement of the Union with the accession of 10 Central and Eastern European countries, the CF beneficiaries significantly increased. In particular, during that programming period, the fund was available to 16 countries, with the number of regions and the affected population more than doubling. It is interesting to note that of the 100 recipient regions, 74 exhibited a level of GDP per capita less than 75\% of the EU average, a significant increase compared to the previous period (essentially, 3 out of 4 beneficiaries were less developed regions). These numbers remained stable for the 2007-2013 period; however, in the 2014-2020 period with the number of eligible countries decreasing, 79 out of the 83 treated regions were classified as less developed. 

It is also worth noting that the enlargement of the EU brought about a significant increase in the budget allocated to the CF. More specifically, during the third programming period (2007-2013), the CF budget more than doubled reaching almost \euro70 billion or 27\% of the overall cohesion policy budget. The 2014-2020 period saw a decrease in the CF budget, which however still accounted for more than 20\% of the total cohesion policy budget. 

This set of evidence highlights the fact that the CF, with its expanding reach over time that covers almost exclusively less developed regions and the increasing budget that aimed to fund environment and transport investment projects --which are crucial for achieving sustainable growth-- is of central importance for ensuring increased cohesion across the EU and merits an in depth analysis of its potential effects. 

\section{Data and Descriptive Statistics}\label{sec:data}
We analyze data from two sources: (i) the Cohesion portal\footnote{\url{https://cohesiondata.ec.europa.eu/Other/Historic-EU-payments-regionalised-and-modelled/tc55-7ysv/about_data}} from which we source data on the EU's Cohesion Fund expenditures (in current prices) and (ii) the Annual Regional Database (ARDECO) of the European Commission\footnote{\url{https://knowledge4policy.ec.europa.eu/territorial/ardeco-database_en}} for macroeconomic data. 

Starting with the Cohesion Fund data, we note that they are generally available for the 1993-2022 period.\footnote{Although the Cohesion Fund was officially launched in 1994, there were some payments in advance towards the eligible countries in 1993.} However, the vast majority of regions became eligible for the CF payments in 2000 (as already mentioned, prior to 2000, only four countries received Cohesion Fund payments). Given that the data are provided in nominal terms, we convert them to 2015 prices using the regional GVA deflator. We note here that prior to using the data for the empirical analysis, we had to account for two issues: first, the fact that the data are not reported under a single NUTS2 classification; and, second, the issue of the time of the recording of Cohesion Fund expenditures. Further details on how we deal with these issues are available in Appendix~\ref{app:data}.   

Figure~\ref{fig:yearly} depicts the evolution of the CF payments over time for all recipient countries, while Figure~\ref{fig:share} depicts the share of disbursements received by the eligible countries. As is evident from the Figure, during the 2007-2013 period there was a significant increase in the CF payments as a means of accommodating the accession of the newest Member States. These increased payments continued up to 2015, due to the well-known delays in the disbursement of EU funds (for more details on this, see \citet{dicharry}). In total, for the period covered by our sample, \euro165.15 billion were disbursed to the recipient countries, with Poland and Spain receiving around 27\% and 17\% of total payments, respectively. 

\begin{figure}[H]
    \centering
    \begin{minipage}{0.48\textwidth}
        \centering
        \caption{\footnotesize{Yearly Cohesion Fund payments}}
        \includegraphics[scale=0.4]{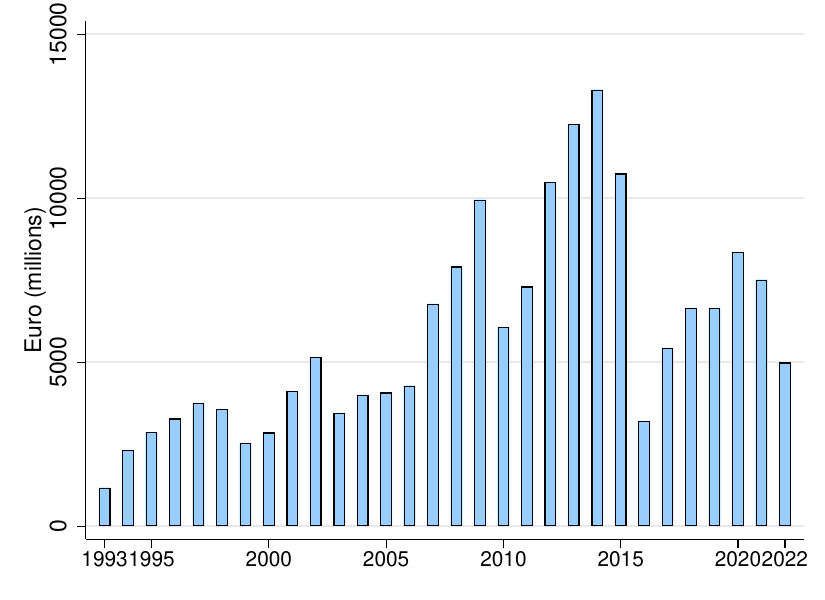}
        \label{fig:yearly}
    \end{minipage}
    \hfill
    \begin{minipage}{0.48\textwidth}
        \centering
        \caption{\footnotesize{Share of Cohesion Fund payments}}
        \includegraphics[scale=0.4]{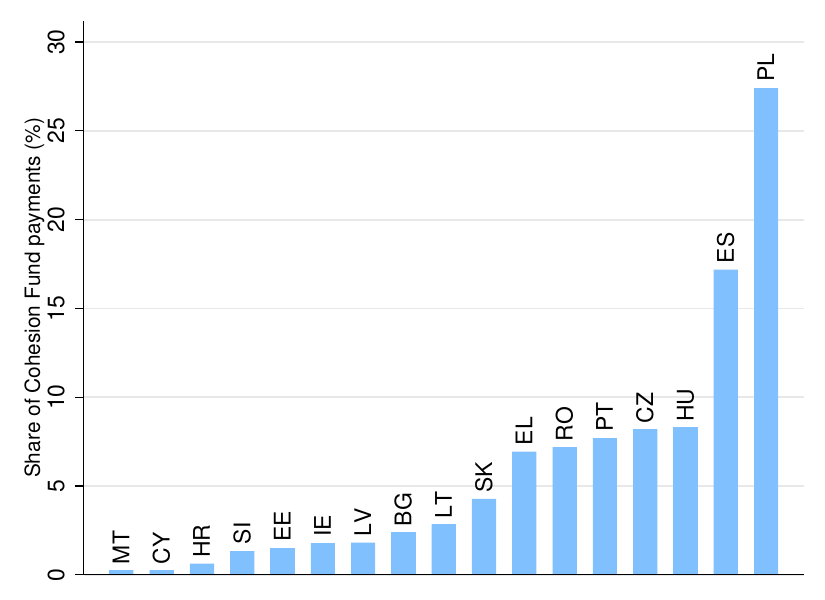} 
        \label{fig:share}
    \end{minipage}
    \caption*{\footnotesize{Note: Figure~\ref{fig:yearly} depicts the total expenditures for the Cohesion Fund by year, while Figure~\ref{fig:share}) presents the average shares of expenditures received by the 17 recipient countries. Note that not all countries received Cohesion Fund payments throughout the sample, e.g., Spain was ineligible for the 2014-2020 period.}}
\end{figure}

Table~\ref{tab:cf_country} provides a disaggregation of the CF disbursements by beneficiary, along with some additional data on the relative importance of these payments in terms of GVA and public investment. The Table covers the 2000-2022 period, which includes the last three programming periods, in order to ensure maximum coverage in terms of recipient countries. We note here that Ireland is excluded from the analysis for two reasons: firstly, Ireland was eligible for the Cohesion Fund for the first two programming periods (1994-1999 and 2000-2006). Secondly, due to the well-documented issues with the Irish National Accounts stemming from the distortionary impact of large multinational enterprises operating within the country it would be increasingly difficult to disentangle the effects of the Cohesion Fund on economic performance.\footnote{For the Ireland's distorted national accounts see \cite{Kostarakosetal2024}} Overall, it is evident from the Table that Poland has received the bulk of the payments over the last 23 years, with a total amount that is more than twice in magnitude compared to the one received by Spain, which ranks second. It is worth highlighting here that Spain was not eligible for funding from the CF during the last programming period (2014-2020).   

Turning to the relative size of the CF payments in terms of GVA, we observe that across beneficiaries the CF ranges from 0.12 to 0.68\%, with a median value of 0.39\%. As such, the contribution of the EU's CF transfers constitute a significant boost of liquidity for the implementation of investment projects. This is further exemplified in the last column of the Table, which depicts the CF transfers as a share of public investment. We observe that, with a median value of 11.7\%, the CF transfers emerge as a significant source of additional funds for the financing of investment, in a complementary fashion to public investment. In particular, in Eastern countries like Lithuania, Latvia, Hungary, Slovakia and Poland, the CF disbursements amounted to at least 15\% of public investment.

\begin{table}[H]
\centering
\caption{Cohesion Fund expenditures, 2000-2022}
\label{tab:cf_country}
\begin{threeparttable}
\begin{tabular}{ccS[table-format=3.2]S[table-format=3.2]S[table-format=3.2]}
\toprule
\centering
Country & Regions & {CF amount} & {CF(\% GVA)} & {CF (\% Gov. Investment)} \\
\midrule
BG & 6 &4206.9& 0.48& 12.46 \\
CY & 1 &459.3& 0.12& 3.9 \\
CZ & 8 &14365.4& 0.42& 11.7  \\
EE & 1 &2634.8& 0.65& 13.3 \\
EL & 13 &8884.4& 0.22& 5.8 \\
ES & 19 &20079.3&0.13 & 3.3 \\
HR & 4 &1110.9& 0.17& 4.1 \\
HU & 8 &14537.8& 0.67& 17.1 \\
LT & 2 &4975.9& 0.68& 15.89 \\
LV & 1 &3169.9& 0.67& 15.13\\
MT & 1 &459.3& 0.26& 8.3 \\
PL & 17 &47988.7& 0.56& 27.1 \\
PT & 7 &9186.6& 0.25& 6.9 \\
RO & 8 &12592.7& 0.39& 14.4 \\
SI & 2 &2348.5& 0.3& 6.5 \\
SK & 4 &7495.9& 0.48& 15.5 \\
\bottomrule
\end{tabular} 
\begin{tablenotes}
\scriptsize
\item \textit{Notes}: The Table covers the 2000-2022 period as in the first programming period (1994-1999) during which the CF was operational only 4 countries were eligible for the funds. 
\item Values are in \euro millions, 2015 prices. Country-level GVA deflators were used.
\item Our proxy for general government investment is the Gross Fixed Capital Formation of the public sector (NACE codes: O to Q) from Eurostat's industry accounts (series: nama\_10\_a64\_p5).
\item The programming periods covered in the Table are 2000-2006, 2007-2013 and 2014-2020. We note that Spain was not eligible for the Cohesion Fund in the 2014-2020.  
\end{tablenotes}
\end{threeparttable}
\end{table}

Our source for macro-related data is ARDECO. We obtain data for real Gross Value Added --GVA-- (series code: SOVGE) and Gross Fixed Capital Formation --GFCF-- (series code: ) at constant 2015 prices. Moreover, we use data for the average annual population at the regional level (series code: SNPTD) and total employment (series code: SNETD). We log-transform the outcome variables, that is GVA per capita and the rate of investment. As such, the empirical results will be interpreted as the percentage change in each variable following the disbursement of of the Cohesion Fund transfers (i.e. after the treatment).



Our initial dataset consists of an unbalanced panel of 242 EU NUTS2 regions belonging to 27 EU Member States — based on the 2021 classification — and spans the 1985–2022 period. We exclude from the sample the so-called outermost regions of France, Portugal, and Spain,\footnote{Specifically, we drop from the sample five French outermost regions, the Azores and Madeira from Portugal, and from Spain the Canary Islands along with the autonomous cities of Ceuta and Melilla.} while we also exclude Ireland for the reasons mentioned above. As such, we end up with a sample of 228 EU NUTS2 regions and 8,249 observations. The sample of treated regions consists of 96 EU NUTS2 regions over the 1993–2022 period, amounting to a total of 2,297 observations and an average of 23.9 years under treatment. These regions belong to 16 countries, the majority of which are the newest Member States of the EU, i.e. the Central and Eastern European countries that entered the Union starting with the 2004 enlargement. Importantly, no region is treated before 1993 — the first year in which Cohesion Fund support becomes available. From that point onward, only 18 regions (located in Greece and Portugal) receive funding continuously for the entire treatment period (1993–2022), while the remaining 78 treated regions receive support for up to 23 years.

\section{Empirical Strategy}\label{sec:empirical}
\subsection{Notation and preliminaries}

Let $y_{it}$ denote the outcome of interest\footnote{In this paper we consider two outcomes of interest; the logarithm of real GVA per capita and the logarithm of Gross Fixed Capital Formation.}, for the $i$-th EU region at year $t$, $i=1,\ldots,N$ and $t=1,\ldots,T$. Throughout the rest of the paper we refer to regions that were eligible for CF payments as the \textit{treated regions}, we call the regions that did not receive this type of funding the \textit{control regions} and we use the term \textit{treatment} when a region receives funding from the CF program. Let $N_{c}$ and $N_{tr} =N-N_{c}$ be the numbers of control and treated units, respectively.  We also denote by $T_i$ the last time period before unit $i$ received the intervention, i.e, $T_i = T$ for control units, while $\mathcal{N}$ denotes the set of all treated regions. 

We work under the framework of potential outcomes (see \citet{holland1986statistics}) to assess the impact of the policy intervention implied by the CF program by relying on observational time-series. For each region and at each time period two potential outcomes are considered; the outcome that region $i$ would have if it was not treated at time $t$, denoted by $y_{it}(0)$ and typically referred as the \textit{potential untreated outcome}, and the outcome that the $i$-th region would have at time $t$ if it was treated, namely the \textit{potential treated outcome} denoted by $y_{it}(1)$. Notice that only one of these outcomes is observed, while the other is known as the \textit{counterfactual} outcome. Specifically, for a treated region $i \in \mathcal{N}$ at time $t > T_i$ we observe $y_{it} = y_{it}(1)$ whereas for $t \leq T_i$ as well as for $i \notin \mathcal{N}$ and for each $t=1,\ldots,T$ we observe $y_{it}=y_{it}(0)$. The causal effect $\tau_{it}$ of the policy intervention is defined in terms of the difference between the observed outcome and the one that region $i$ would have if it was not treated at time $t$, i.e., the counterfactual, so that for treated regions and $t>T_i$:
\begin{align}
\label{eq:tau_def}
    \tau_{it} &= y_{it}(1) - y_{it}(0) \nonumber \\
    & =y_{it} - y_{it}(0).
\end{align}
Finally, let $D_{it}$ denote the treatment indicator, $D_{it} =1$ if region $i$ has received the intervention at time $t$ and $D_{it} =0$ otherwise.

\subsection{Causal inference by using matrix completion}

To assess the effect of the policy intervention implied by the CF program, we follow recent directions in the literature and particularly the factor-augmented approach \citep{xu2023causal}, where unobserved confounders are modeled through interactive fixed effects \citep{bai2009panel}. Specifically, we employ the matrix completion method \citep{athey2021matrix,borusyak2024revisiting}, considering the unobserved potential untreated outcomes $y_{it}(0)$, $i \in \mathcal{N}$ and $t > T_i$, as missing values to be estimated as $\widehat{y}_{it}(0)$. This allows us to estimate the causal effect $\tau_{it}$ as 
\begin{equation}
\label{eq:tauhat}
    \widehat{\tau}_{it} = y_{it} - \widehat{y}_{it}(0).
\end{equation}

In particular, we assume for each $i=1,\ldots,N$ and $t=1,\ldots,T$ the following factor model for both observed and potential untreated outcomes becomes 
\begin{equation}
    \begin{aligned}
        y_{it} &= \tau_{it}D_{it} + X_{it}^\top\beta  + \lambda_i^\top f_t + \epsilon_{it} \\
        &= X_{it}^\top\beta + \lambda_i^\top f_t + \epsilon_{it}
    \end{aligned}
    \label{eq:factor_model}
\end{equation}
where $X_{it}$ is a $p \times 1$ vector of observed covariates and $\beta$ the corresponding coefficient vector.\footnote{Following \citet{gobillon2016regional} and \citet{xu2017generalized}, among others, we assume that $\beta$ is constant across regions and time. This is justified by the use of interactive fixed effects, which flexibly capture complex region-time heterogeneity. Allowing $\beta$ to vary across $i$ and $t$ introduces estimation complexity and risks confounding observed and unobserved components; see also \citet{athey2021matrix}.} 
The covariates $X_{it}$ account for observed sources of heterogeneity and improve identification under the standard conditional independence assumption 
\[
\{Y_{it}(1), Y_{it}(0)\} \perp D_{it} \mid X_{it}, \lambda_i, f_t.
\]
In practice, this condition is typically satisfied when $X_{it}$ captures pre-treatment or predetermined characteristics. Conditioning on post-treatment variables is not required for identification and can change the estimand by blocking causal channels. The term $f_t$ is a $K \times 1$ vector of unobserved common factors, and $\lambda_i$ is a $K \times 1$ vector of region-specific loadings; together, they flexibly capture unobserved time-varying heterogeneity and cross-sectional dependence. The second line of \eqref{eq:factor_model} emphasizes that for potential untreated outcomes we have $D_{it}=0$ for all $i$ and $t$, i.e., \eqref{eq:factor_model} also describes the data-generating process of untreated outcomes.

Moreover, the model in \eqref{eq:factor_model} generalizes the standard Two-Way Fixed Effects (TWFE) specification, which is widely used for causal estimation under the Difference-in-Differences (DiD) framework; see Appendix~\ref{app:technical} for more details. Under its identifying assumptions, the TWFE estimator recovers the parameter $\tau = \mathbb{E}[\tau_{it} \mid D_{it}=1, X_{it}]$, the average treatment effect on the treated (ATT), but it cannot capture heterogeneous, unit- and time-specific effects $\tau_{it}$. Estimation based on the TWFE model relies on several strong assumptions: (i) \emph{additive separability}, requiring unobserved heterogeneity to be captured by additive unit and time fixed effects; (ii) \emph{parallel trends}, which assumes treated and control units would evolve similarly absent treatment; (iii) \emph{no anticipation}, excluding selection on future outcomes; and (iv) \emph{homogeneous treatment effects}, implying constant effects across units and time.

The specification in \eqref{eq:factor_model} mitigates reliance on most of these assumptions. First, the \emph{parallel trends} condition is relaxed because unobserved time-varying confounders that could violate it are captured by the interactive fixed effects term $\lambda_i^\top f_t$. The inclusion of observed covariates $X_{it}$ further helps account for observable differences between treated and control units, improving balance and reducing residual variance, provided these covariates are predetermined and not themselves affected by treatment. The latent factors $f_t$ represent unobserved time-varying shocks common to all regions, while $\lambda_i$ allows these shocks to affect regions heterogeneously. This richer structure captures complex forms of cross-sectional dependence that cannot be addressed by additive fixed effects alone, thereby mitigating the \emph{additive separability} restriction. Finally, allowing treatment effects to vary across both units and time directly relaxes the \emph{homogeneous treatment effects} assumption — a particularly important feature in the context of regional policy evaluation, where differences in economic structure, institutional capacity, or spatial linkages are expected to generate heterogeneous responses to policy interventions.

Importantly, our approach enables the estimation of heterogeneous causal effects $\tau_{it}$ across units and time. We summarize these effects via a time-specific $ATT_t$ (commonly called an event-study estimate), as
\begin{equation}
    \label{eq:ATT_t}
    \widehat{ATT}_t = \frac{1}{|N_{tr,t}|} \sum_{i \in \mathcal{N}} \widehat{\tau}_{it},
\end{equation}
where $N_{tr,t} = \{ i : T_i \leq t \}$ is the set of units treated by time $t$. Overall, the identification of causal effects provided by the factor model in \eqref{eq:factor_model} relies on two main assumptions: (i) \emph{strict exogeneity of treatment}, i.e., treatment is uncorrelated with the idiosyncratic error term after controlling for both the latent factors and the observed covariates $X_{it}$, thus ruling out endogenous selection; and (ii) \emph{no interference between units}, in the sense that the potential outcomes of a given unit should not depend directly on the treatment assignment of other units. This second condition is partly relaxed in practice by the presence of interactive fixed effects, which can accommodate correlated unobserved shocks such as trade linkages or coordinated policy responses. \footnote{While interactive fixed effects can accommodate latent cross-sectional dependence, they do not fully address violations of the no-interference assumption. Explicit modeling of spillovers in panel causal inference remains challenging and requires additional structure or detailed network information often unavailable in regional policy settings. We therefore follow the state-of-the-art approach for imputing counterfactual outcomes by combining interactive fixed effects with matrix completion, building on the foundational work of \cite{gobillon2016regional} and recent advances in low-rank causal imputation \citep[e.g.,][]{athey2021matrix}.}

\subsubsection{Estimation of causal effects}\label{sec:estimation}

To operationalize the model in \eqref{eq:factor_model}, we cast the estimation of counterfactual untreated outcomes $y_{it}(0)$ as a matrix completion problem. As shown in \citet{athey2021matrix}, the interactive fixed effects structure can be equivalently expressed at the matrix level by decomposing the full outcome matrix $Y$ into a low-rank matrix capturing latent effects, a covariate component, and noise. Specifically, the estimation model becomes:
\begin{equation}
\label{eq:model_with_covariates}
Y = L + X \beta + E,
\end{equation}
which corresponds to a matrix reformulation of \eqref{eq:factor_model}, where the latent factor term $\lambda_i^\top f_t$ across $i$ and $t$ is compactly represented by the low-rank matrix $L$; $L_{it} = \lambda_i^\top f_t$. The matrix $X$ stacks the covariates $X_{it}$ appropriately across units and time. Thus, \eqref{eq:model_with_covariates} can be viewed as the global form of the unit-time model in \eqref{eq:factor_model}, and provides a natural starting point for matrix completion techniques to estimate the missing entries corresponding to $y_{it}(0)$.

In particular, we observe a subset of entries in $Y$, indexed by $\mathcal{O} \subset \{1, \ldots, N\} \times \{1, \ldots, T\}$, and estimate $(L, \beta)$ by solving the following regularized convex optimization problem
\begin{equation}
\label{eq:min_covariates}
(\hat{L}, \hat{\beta}) = \arg\min_{L, \beta} \left\{ \|P_\mathcal{O}(Y - L - X \beta)\|_F^2 + \lambda \|L\|_* \right\},
\end{equation}
where $P_{\mathcal{O}}(\cdot)$ is the projection operator that retains entries in $\mathcal{O}$ and zeros out the rest $\|\cdot\|_F$ denotes the Frobenius norm, $\|\cdot\|_*$ denotes the nuclear norm (sum of singular values), $\lambda > 0$ is a tuning parameter that controls the rank of $\hat{L}$.

The optimization problem in \eqref{eq:min_covariates} is solved via an iterative procedure, alternating between (i) estimating $\hat{\beta}$ using least squares on the residuals $Y - L$ over observed entries and (ii) estimating $\hat{L}$ using a matrix completion algorithm (e.g., soft-impute) on the residuals $Y - X\hat{\beta}$. This procedure continues until convergence ; see in the Appendix {\color{red} XX} for more details. Once $\hat{L}$ and $\hat{\beta}$ are obtained, the counterfactual untreated outcomes are imputed as
\begin{equation*}
\hat{y}_{it}(0) = \hat{L}_{it} + X_{it}^\top \hat{\beta}, \quad \text{for } i \in \mathcal{N}, \, t > T_i.
\end{equation*}
Hence, the estimated treatment effects for treated units in post-treatment periods are
\begin{equation*}
\hat{\tau}_{it} = y_{it} - \hat{y}_{it}(0) = y_{it} - \hat{L}_{it} - X_{it}^\top \hat{\beta}.
\end{equation*}
This approach separates latent structure from observed heterogeneity, improving both interpretability and predictive accuracy when covariates are informative.

\section{Results}\label{sec:results}
In this section we present the empirical estimates of the impact of Cohesion Fund payments on regional economic performance. In particular, we estimate region- and time-specific causal effects using the model specified in equation~\eqref{eq:factor_model}, including the employment-to-population (EMP) ratio, $X_{it}$, as a time-varying covariate. The EMP ratio is computed as the total number of employed individuals divided by the working-age population in each region at each time point. This variable is an important confounder to control for, as it is strongly correlated with GVA per capita and serves as a proxy for the strength of the regional labor market. Regions with higher EMP ratios typically exhibit stronger economic performance. Notably, treated regions in our sample tend to have significantly lower EMP ratios relative to the EU average, underscoring the need to adjust for this imbalance in the analysis. Following a similar argument, the employment-to-population ratio is included as a covariate in the estimation of the investment effects of the Cohesion Fund transfers. Specifically, as the employment-to-population ratio is known to exhibit a cyclical pattern and can serve as a proxy of the business cycle, its inclusion in the specification allows for a better identification of the variation in the outcome variable (in the case, the rate of investment).\footnote{The estimation is conducted by using the \texttt{gsynth} library \citep{gsynth} in the statistical software R \citep{citeR}.} Note that EMP is the only covariate available at the regional level for the whole timeframe available. This fact underscores the importance of accounting for unobserved councounders by use of interactive fixed effects, as in \ref{eq:factor_model}. Without allowing for these unobserved factors, the model at a regional level could be highly misspecified due to the lack of regional data.  

The rest of the section is organized as follows: firstly, in section~\ref{sec:ATT} we present the average treatment effect on the treated. Next, in section~\ref{sec:Het}, we delve into the heterogeneous nature of the effects; specifically, section~\ref{sec:Distribution} examines the time-varying distribution of the effects, section~\ref{sec:Maps} focuses on the spatial patterns that emerge and section~\ref{sec:Income} examines the relationship between the income level of the regions and the estimated effects. Section~\ref{sec:Gini} examines the evolution of regional income heterogeneity over time by means of the Gini coefficient. Lastly, section~\ref{sec:NLR} provides some graphical evidence on the non-linear nature of the relationship between funding intensity and its GVA effects. 

\subsection{Average Treatment Effect}\label{sec:ATT}

\begin{figure}[H]
    \centering
    \begin{minipage}{0.48\textwidth}
        \centering
        \caption{\footnotesize{$\widehat{ATT}_t$ for GVA}}
        \includegraphics[scale=0.35]{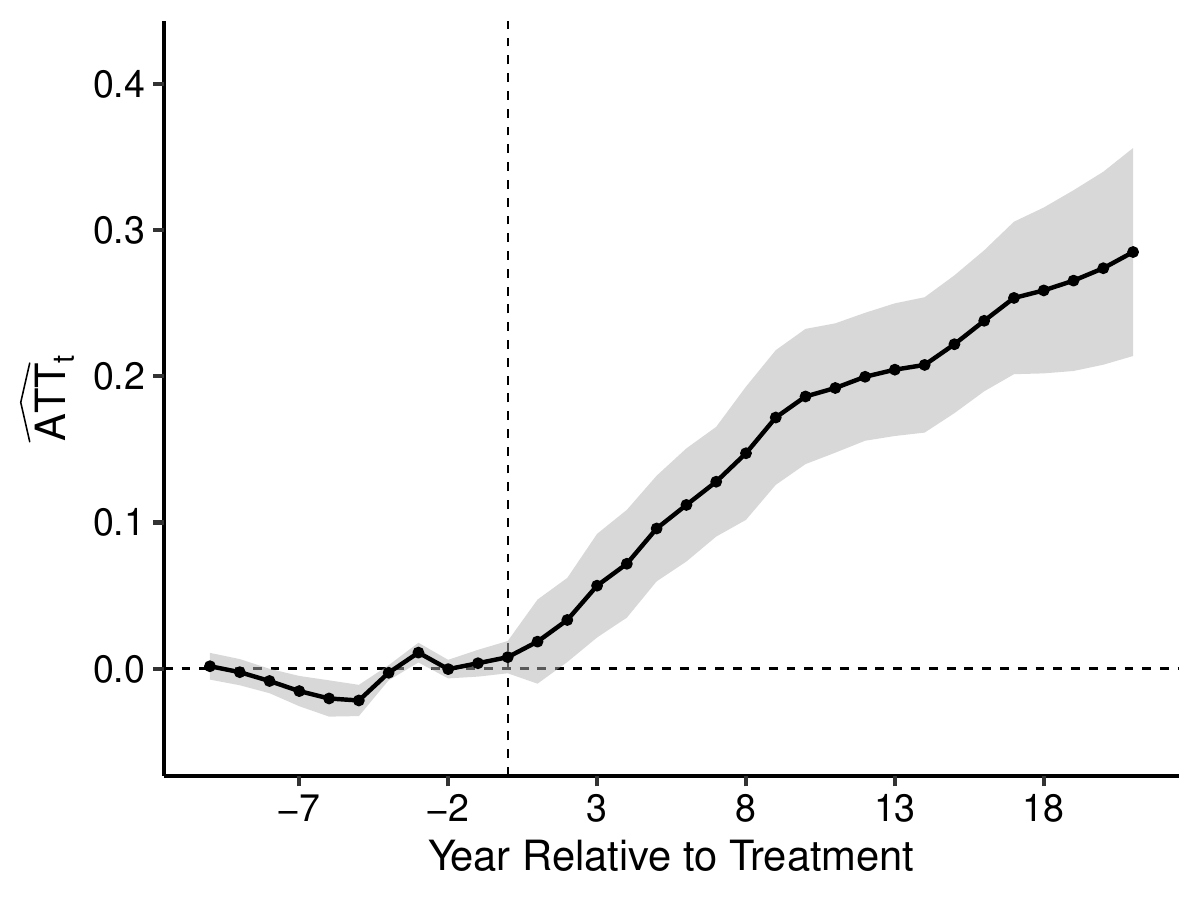}
        \label{fig:ATT1}
    \end{minipage}
    \hfill
    \begin{minipage}{0.48\textwidth}
        \centering
        \caption{\footnotesize{$\widehat{ATT}_t$ for GFCF}}
        \includegraphics[scale=0.35]{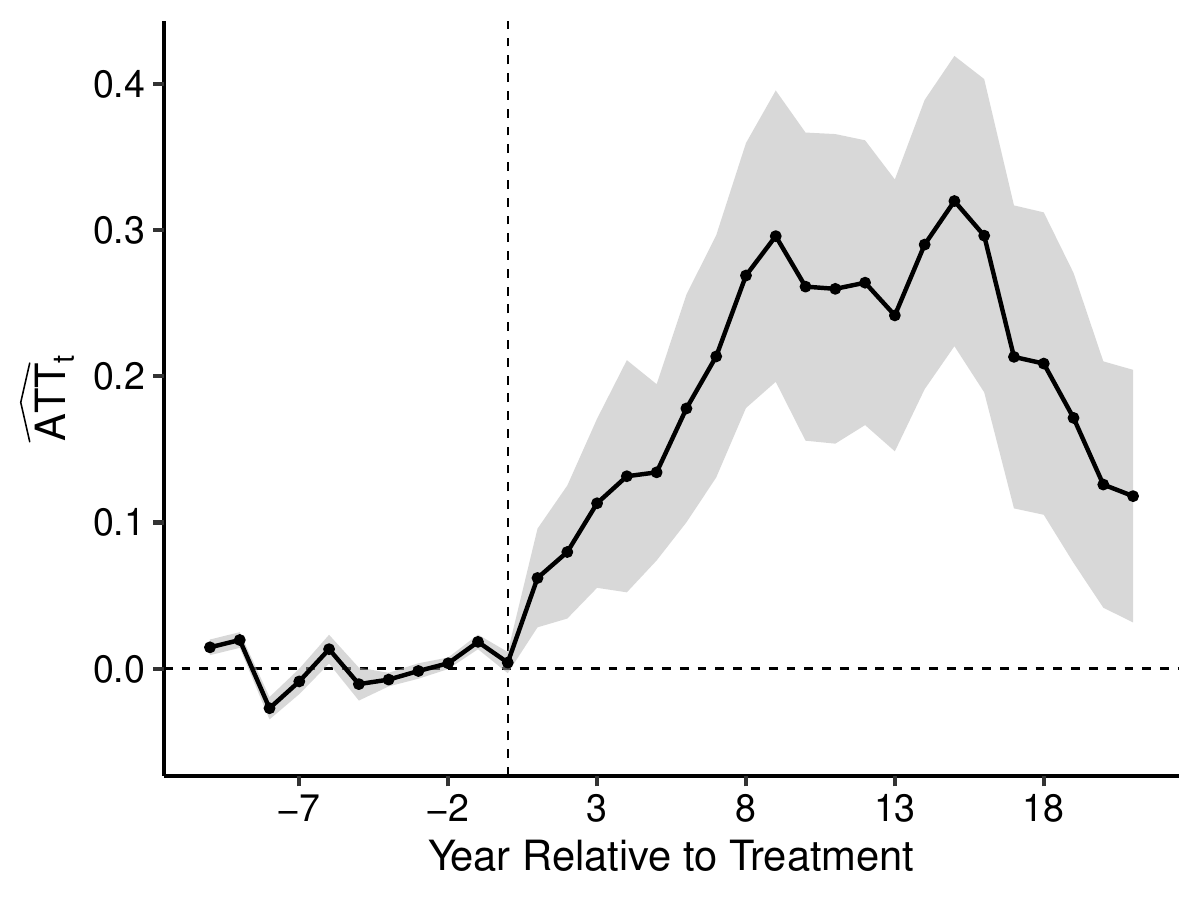} 
        \label{fig:ATT2}
    \end{minipage}
    \caption*{\footnotesize{Note: Average Treatment Effect on the Treated ($ATT$) over time (black solid line) and $95\%$ confidence intervals (gray shaded area) for the log of GVA per capita (left panel) and the log of GFCF/GVA (right panel).}}
\end{figure}

We start by presenting the average treatment effect across all regions as estimated in section \ref{sec:estimation}; specifically, see equation \eqref{eq:ATT_t}. Figures~\ref{fig:ATT1} and ~\ref{fig:ATT2} present the average treatment effect on the y-axis against years relative to treatment for GVA per capita and Gross Fixed Capital Formation (GFCF), respectively. The vertical red line that crosses the x-axis in year 0 represents the initial year in which each region received CF payments. Note that this year might differ for each region as EU countries became eligible for the CF payments at different years and, as such, the regions in our sample received payments at different points in time. Thus, the treatment effect is shown by the black line for each year that follows the zero year. The grey shaded area represents the upper and lower bound of the $95\%$ confidence interval calculated using a parametric bootstrap technique as in \cite{xu2017generalized} (see Appendix~\ref{app:technical} for more details).  

\noindent Figures~\ref{fig:ATT1} and ~\ref{fig:ATT2} show that on average CF payments have a persistent positive effect on both the GVA and GFCF of a region that receives them (treated) compared to the counterfactual scenario where these funds are not provided by the EU.\footnote{Figure~\ref{fig:obs_ct} in Appendix~\ref{app:tables} plots both the observed and the respective estimated counterfactual average GVA in logs as well as in growth rates. Their difference is the average treatment effect presented in Figure~\ref{fig:ATT1}. Additionally, Figure~\ref{fig:obs_ct_growth} depicts the same series for the growth rate of GVA.}  

The average treatment effects depicted on the y-axis of Figure~\ref{fig:ATT1} indicate that within the first seven years under treatment, Cohesion Fund transfers are associated with an increase of almost 17\% in GVA per capita. By the end of three consecutive seven-years transfer cycles (i.e., after 21 years after inclusion in the CF program), GVA per capita is, on average, 30\% higher than it would have been in the absence of CF (in log scale\footnote{Suppose a hypothetical treated average EU region $i$
that received treatment in period $0$. If $\widehat{ATT}_t = 0.3$, then
$\log\!\left(\frac{\text{GVA}_{it}}{\widehat{\text{GVA}}_{it}(0)}\right) = 0.3$,
so $\frac{\text{GVA}_{it}}{\widehat{\text{GVA}}_{it}(0)} = e^{0.3} \approx 1.35$,
i.e.\ $\text{GVA}_{it} \approx 1.35 \times \widehat{\text{GVA}}_{it}(0)$, where
$\widehat{\text{GVA}}_{it}(0)$ is GVA in period $t$ under the counterfactual
scenario that region $i$ had not received CF.}
As such, the majority of the income increase appears to materialize in the early years of the implementation of the policy programme, while further gains occur but are relatively smaller. This is an indication that the CF programme effects are persistent but quite frontloaded. In terms of the average impact on investment (GFCF), as depicted in Figure~\ref{fig:ATT2}, we observe again a persistent positive impact that appears to be frontloaded;  however, the average impact is both larger in magnitude and exhibits a more volatile path over time. Specifically, within the first seven years under treatment, the share of investment is almost 45\% higher. Following this, the impact plateaus and exhibits a slight decline; nonetheless, by the end of three consecutive seven-year periods, the share of investment is almost 40\% higher.\\
\par
Using the estimates presented in Figure~\ref{fig:ATT1}, we compute the net benefit for the average treated region due to its inclusion in the CF expressed in levels of GVA per capita. Specifically, GVA per capita increases by \euro 815 more that it would under the counterfactual scenario of no Cohesion Fund support.\footnote{Specifically, for the hypothetical EU27 treated region, GVA per capita rises by \euro2168 over the first seven years of treatment\textemdash from \euro 9,064 in the first year to \euro 11,232 in year seven. Under the counterfactual scenario, the corresponding increase is \euro1353 \textemdash from \euro8618 to \euro9971 in year seven.} Similarly, on the 27th year under treatment the Cohesion Funds transfers lead to an increase in GVA per capita equal to 31.3\%, while for the entire period the average increase is 28.5\%. This implies that the hypothetical treated region experiences an increase in income that is \euro2961 larger than it would have been under the counterfactual scenario. 
\noindent Following a similar calculation, a hypothetical treated region experiences an increase in investment per capita that is \euro395 and \euro680 more than it would have been under the counterfactual scenario after 7 and 21 years of treatment, respectively.\footnote{Figures \ref{fig:GVA_MCNNvsTWFE} and \ref{fig:GFCF_MCNNvsTWFE} in Appendix \ref{COMP_TWFE} show that the estimated average treatment effect of this section is quantitatively very close to the analogous estimate obtained from the traditional Two-Way Fixed Effects approach.}

\subsection{Heterogeneous effects}\label{sec:Het}

\subsubsection{Quantitative analysis of the heterogeneous effects}\label{sec:Distribution}

The previous section focused on the average treatment effect across all regions that have received CF payments. However, a key advantage of our empirical strategy is that it allows us to explore the heterogeneous effects of CF across the different regions in our sample. 

\begin{figure}[H]
    \centering
    \centering
    \caption{\footnotesize{Distribution of GVA effects}}
    \includegraphics[scale=0.5]{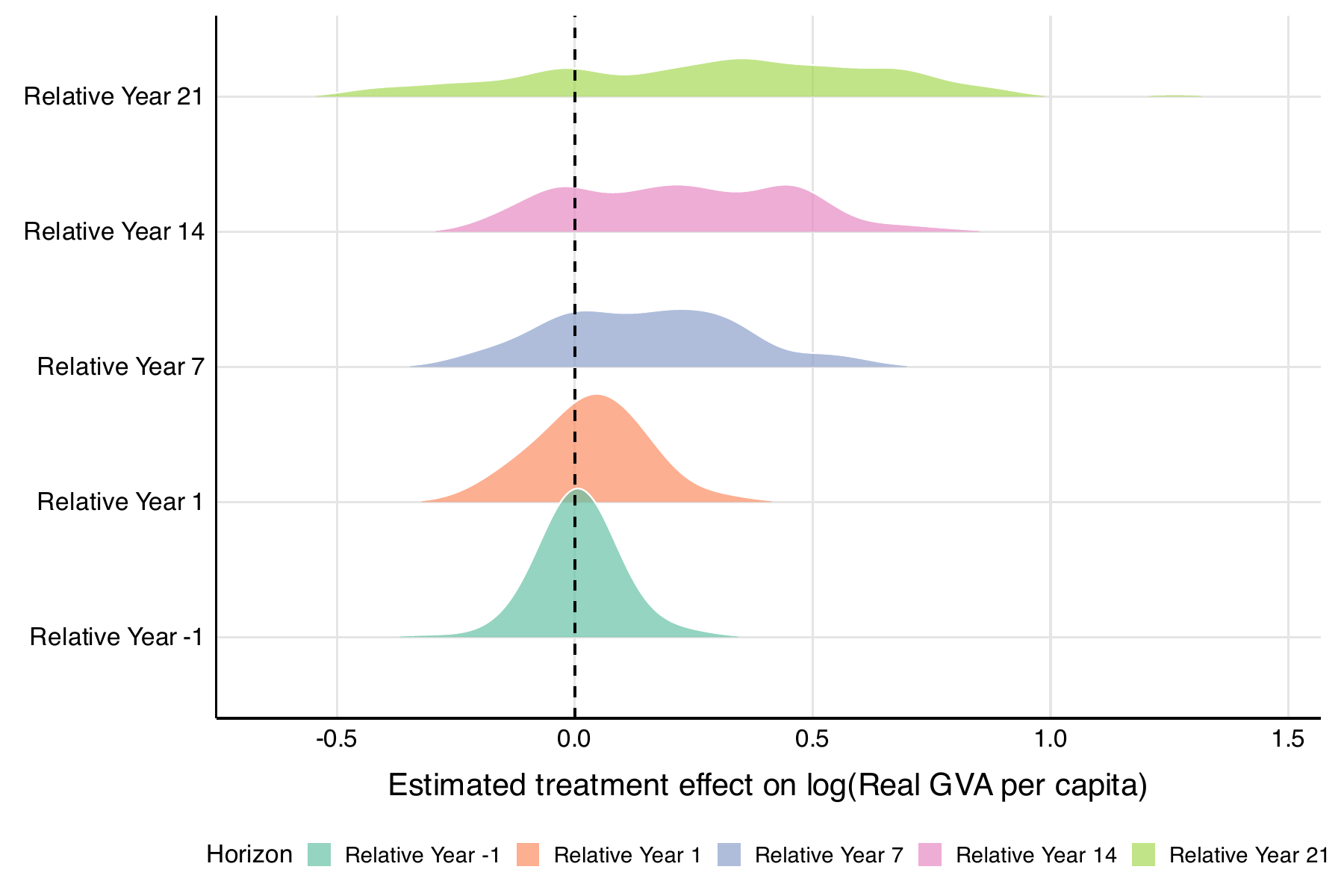}
    \label{fig:DISTRIBUTION1}
    \end{figure}

Figure~\ref{fig:DISTRIBUTION1} illustrates the distribution of treatment effects on log real GVA per capita across treated regions at selected years relative to treatment. The kernel density estimates indicate that the distribution is centered near zero one year prior to treatment, as expected (see the bottom panel of Figure~\ref{fig:DISTRIBUTION1}). Post-treatment, the distribution gradually shifts rightward, with increasing mass above zero at years 7, 14, and 21, suggesting that the policy intervention generated increasingly positive economic effects over time. The widening of the distribution at longer horizons also indicates growing heterogeneity in regional responses.

\begin{figure}[H]
    \centering
    \caption{\footnotesize{Distribution of GFCF effects}}
    \includegraphics[scale=0.5]{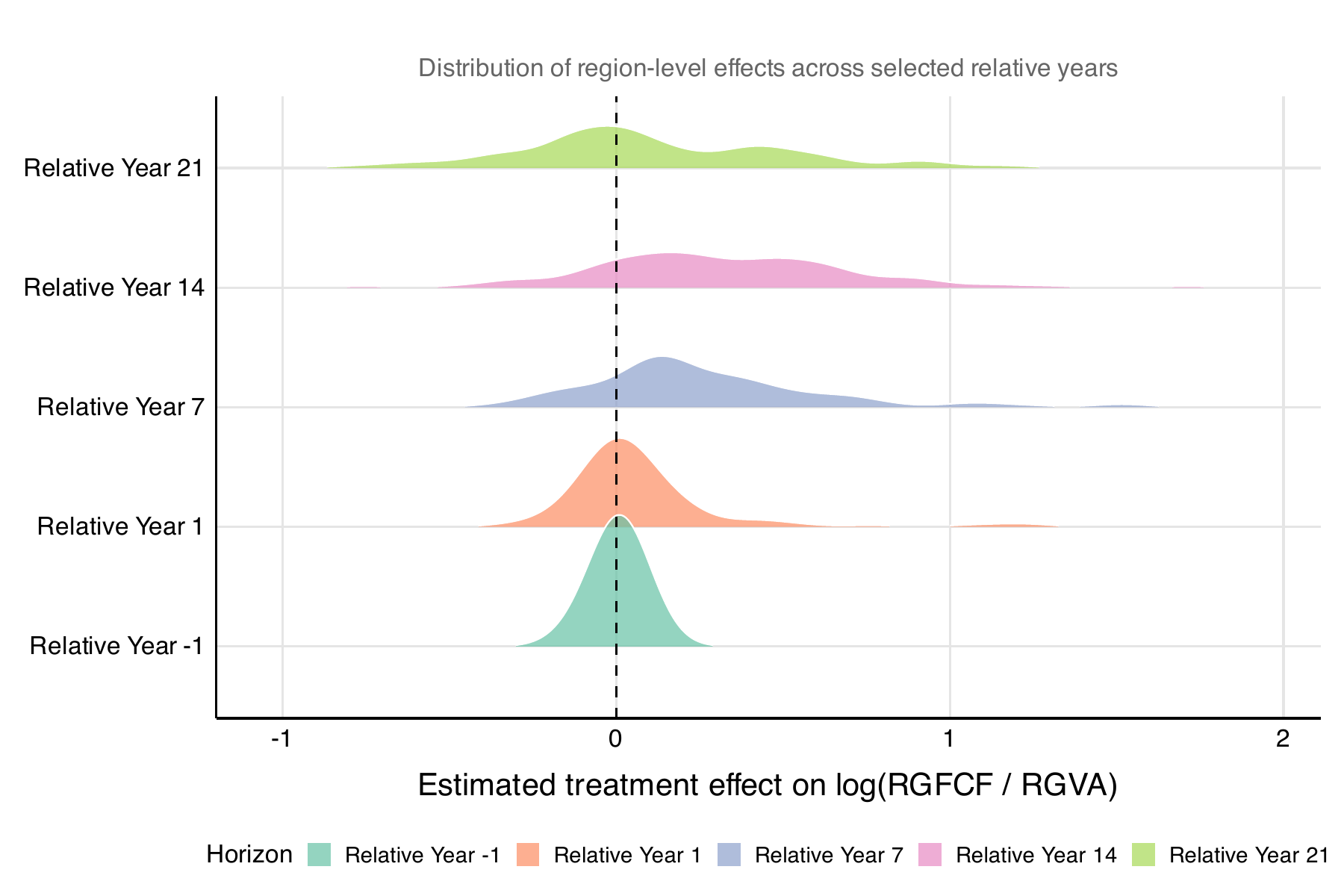}
    \label{fig:DISTRIBUTION2}
\end{figure}

Figure~\ref{fig:DISTRIBUTION2} shows analogous results for log real gross fixed capital formation (GFCF). Again, the pre-treatment distribution is centered near zero, while post-treatment distributions, particularly at years 14 and 21, shift notably to the right and exhibit longer right tails. This pattern suggests that investment responses to the treatment are not only positive on average but also more skewed, indicating that some regions experienced particularly large in magnitude increases in investment.

The analysis in this section indicates that focusing solely on average treatment effects (as in section \ref{sec:ATT}) obscures substantial heterogeneity in the effects of EU Cohesion Funds (CF). Recognizing this heterogeneity can provide valuable insights for EU policymakers when determining the optimal allocation of EU resources. \footnote{ Indicatively, the observed heterogeneity could be related to the intensity of the treatment, i.e. the size of the Cohesion Fund expenditures relative to GVA. If a maximum desired intensity is identified, this provides flexibility to reallocate the resources and achieve better outcomes.} This is a novelty of our approach, as most existing papers usually limit their analysis to the estimation of an average effect across regions, ignoring any potential heterogeneity. We touch upon this issue in more detail in section~\ref{sec:NLR}.


Table~\ref{merged_stats} presents summary statistics for the time-varying distributions of the CF effects on regional gross value added (GVA) and gross fixed capital formation (GFCF), corresponding to the distributions shown in Figures~\ref{fig:DISTRIBUTION1} and ~\ref{fig:DISTRIBUTION2}, respectively. Starting with the effects on GVA, in the top panel of the Table, the results indicate that participation in the CF program tends to have a predominantly positive impact on GVA for the majority of treated regions (see first panel). Specifically, during the first seven-year cycle, $75.32\%$ of treated regions experience positive effects; this increases to $80.2\%$ after 14 years and marginally declines to $78.2\%$ after 21 years. On average, $80\%$ of the treated regions benefit from the CF in terms of GVA per capita. The skewness of the distribution (second row) shows a positively skewed effect in the first and second cycles, indicating that while most regions experience modest gains, a smaller number benefit disproportionately. However, skewness declines by the third cycle, suggesting that the most extreme positive effects become less frequent over time. Kurtosis follows a similar trajectory, although relatively moderate in earlier cycles, the average value rises to 1.99, implying a distribution with heavier tails—i.e., some regions experience particularly large or small effects relative to the mean.

The second panel reports the distributional characteristics of CF effects on GFCF (i.e., from Figure \ref{fig:DISTRIBUTION2}). Although most of the patterns observed for GVA also hold here, notable differences emerge. The impact on regional investment is significantly larger in magnitude and affects a greater proportion of treated regions. Nearly $89\%$ of regions experience a positive effect on investment in the first seven years—substantially higher than the corresponding figure for GVA. This share increases to almost $91\%$ after 14 years and then declines to $74\%$ after 21 years, with an overall average of $88.5\%$. The distribution of the investment effects also differs in shape. Skewness is higher in the early years and even turns slightly negative by the third cycle, while kurtosis consistently exceeds that of the GVA distribution. This suggests a more pronounced right-skewed shape early on, with fatter tails, meaning that although most regions benefit, some see particularly strong gains in investment. These findings align with the core objective of the CF program, which is to stimulate investment and infrastructure development in less developed EU regions. The frontloaded nature of these effects on GFCF implies that capital formation is an early outcome of CF participation. Moreover, this investment may serve as a transmission channel through which the observed longer-term gains in GVA are realized.

\begin{table}[H]
\centering
\begin{threeparttable}
\caption{Summary statistics for the distribution of the regional effects}
\label{merged_stats}
\begin{tabular}{lcccc}
  \hline
 & 7 years & 14 years & 21 years & Average\tnote{$\dagger$} \\ 
  \hline
  \textbf{GVA per capita} & & & & \\
  $\%$ of regions with $\widehat{\tau}_i > 0$ & 75.32 & 80.2 & 78.2 & 80.2 \\
  Skewness & 0.17 & 0.18 & -0.16 & 0.04 \\ 
  Kurtosis & 2.74 & 1.87 & 2.19 & 1.99 \\ 
  \addlinespace
  \textbf{GFCF} & & & & \\
  $\%$ of regions with $\widehat{\tau}_i > 0$ & 88.5 & 90.6 & 73.9 & 88.5 \\
  Skewness & 0.97 & 0.39 & -0.04 & 0.84 \\ 
  Kurtosis & 4.31 & 2.56 & 2.25 & 3.78 \\ 
  \hline
\end{tabular}
\begin{tablenotes}
\scriptsize
\item \textit{Notes}: In the 21st year under treatment the sample consists of 92 regions, as the regions in Croatia are not treated. 
\item \tnote{$\dagger$} This column depicts the average based on the entire post-treatment period.
\end{tablenotes}
\end{threeparttable}
\end{table}

\subsubsection{Geographical analysis of the heterogeneous level effects}\label{sec:Maps}

\begin{figure}[H]
    \centering
    \caption{Spatial distribution of GVA per capita effects}
    \begin{minipage}{.3\textwidth}
        \centering
        \includegraphics[width=\linewidth]{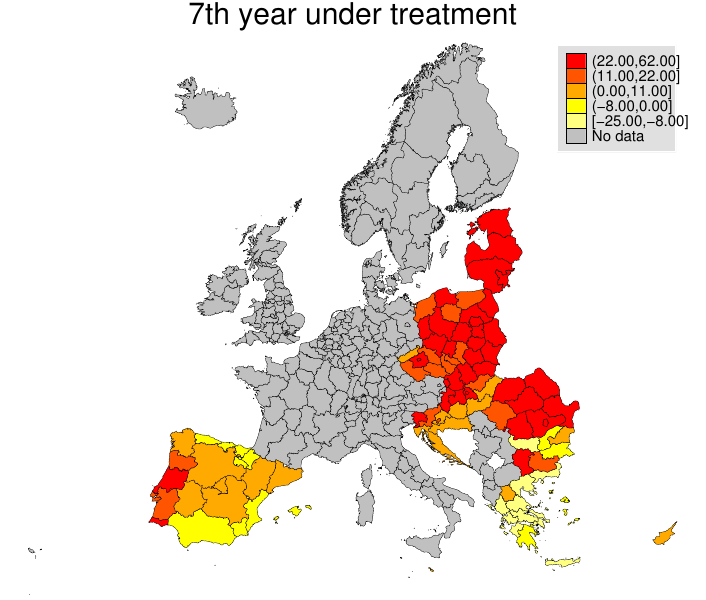}
    \end{minipage}%
    \begin{minipage}{0.3\textwidth}
        \centering
        \includegraphics[width=\linewidth]{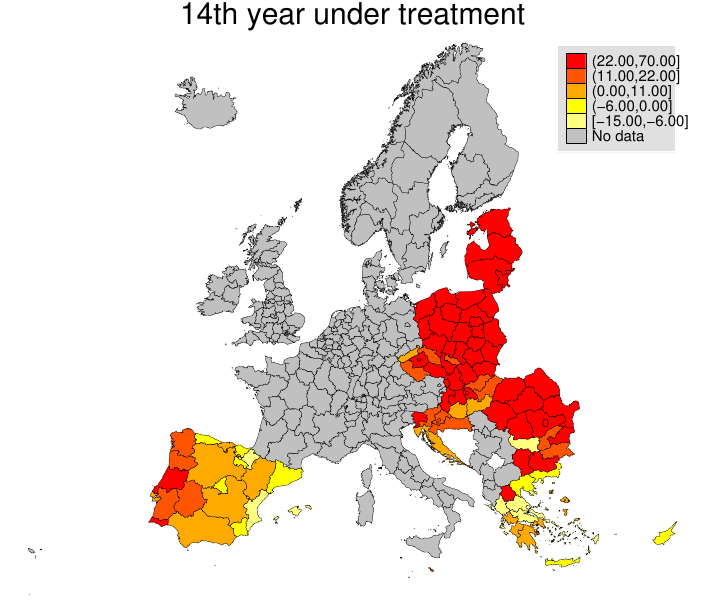}
    \end{minipage}
    \begin{minipage}{0.3\textwidth}
        \centering
        \includegraphics[width=\linewidth]{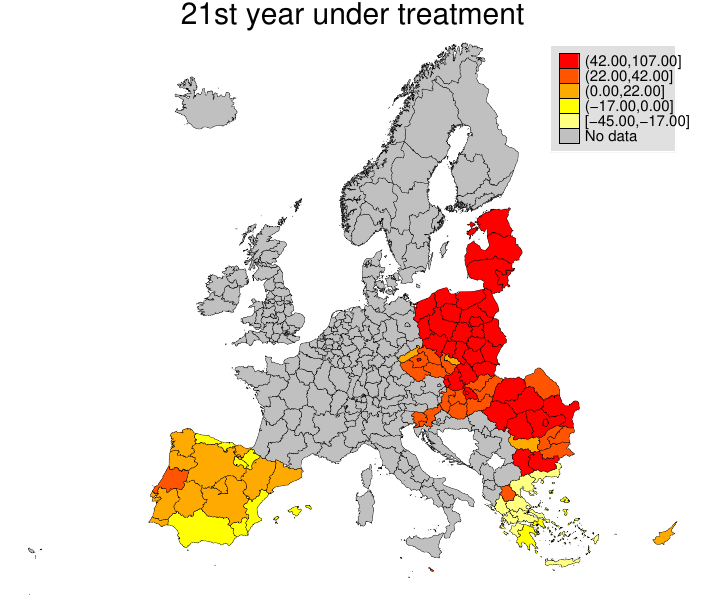}
    \end{minipage}
     \label{fig:map1}
\end{figure}

This section illustrates the geographical dispersion of EU Cohesion Fund's policy impact across Europe. To do this, we categorize regions into five percentiles based on their effects on GVA and GFCF ranging from the lowest (light colors\textemdash yellow) to the highest (dark colors\textemdash red). Figure~\ref{fig:map1} maps the GVA effect across geographical regions.

A distinct spatial pattern emerges from the visual inspection of the map. The smallest effects of the Cohesion Fund, represented by lighter colors (yellow), are concentrated in Southern Europe. In contrast, the largest effects are observed in Eastern Europe (red), while gray-shaded regions, corresponding to Northern Europe, were not eligible for CF support. 

This pattern reveals a new  geographical division within Europe. In this case, Southern Europe, which consists of relatively wealthier regions, experiences weaker CF effects compared to Eastern Europe, where poorer regions benefit the most from such a policy. As such, the newest Member States that gained their EU status post-2000 are the ones that seem to reap the benefits of the Cohesion Fund assistance.

\begin{figure}[H]
    \centering
    \caption{Spatial distribution of Investement effects}
    \begin{minipage}{.3\textwidth}
        \centering
        \includegraphics[width=\linewidth]{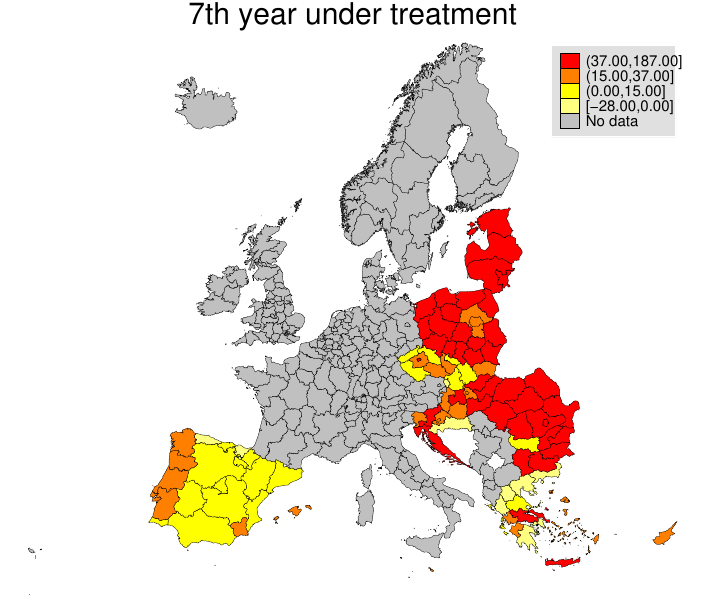}
    \end{minipage}%
    \begin{minipage}{0.3\textwidth}
        \centering
        \includegraphics[width=\linewidth]{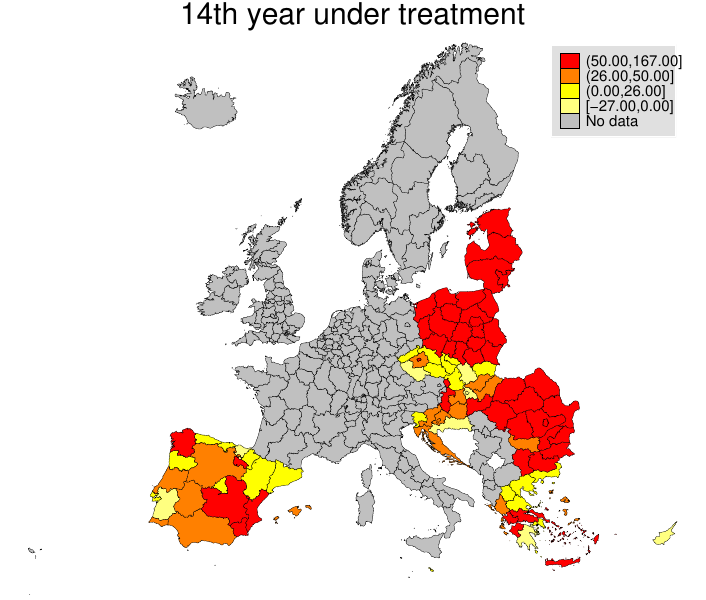}
    \end{minipage}
    \begin{minipage}{0.3\textwidth}
        \centering
        \includegraphics[width=\linewidth]{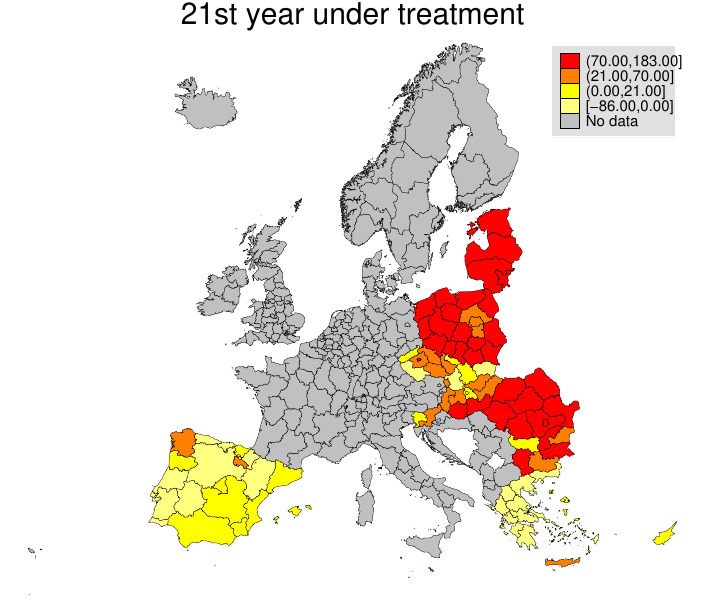}
    \end{minipage}
     \label{fig:map2}
\end{figure}

A similar pattern emerges when focusing on the effects on investment. As can be seen in Figure~\ref{fig:map2} the largest effects are again observed in Eastern Europe, while Southern Europe experiences milder impacts. 

\subsubsection{Heterogeneous impact of the Cohesion Fund by income group}\label{sec:Income}

\begin{figure}[H]
    \centering
        \caption{\footnotesize{$\widehat{ATT}$ by quintile based on pre-treatment GVA per capita distribution.}}
    \includegraphics[scale=0.5]{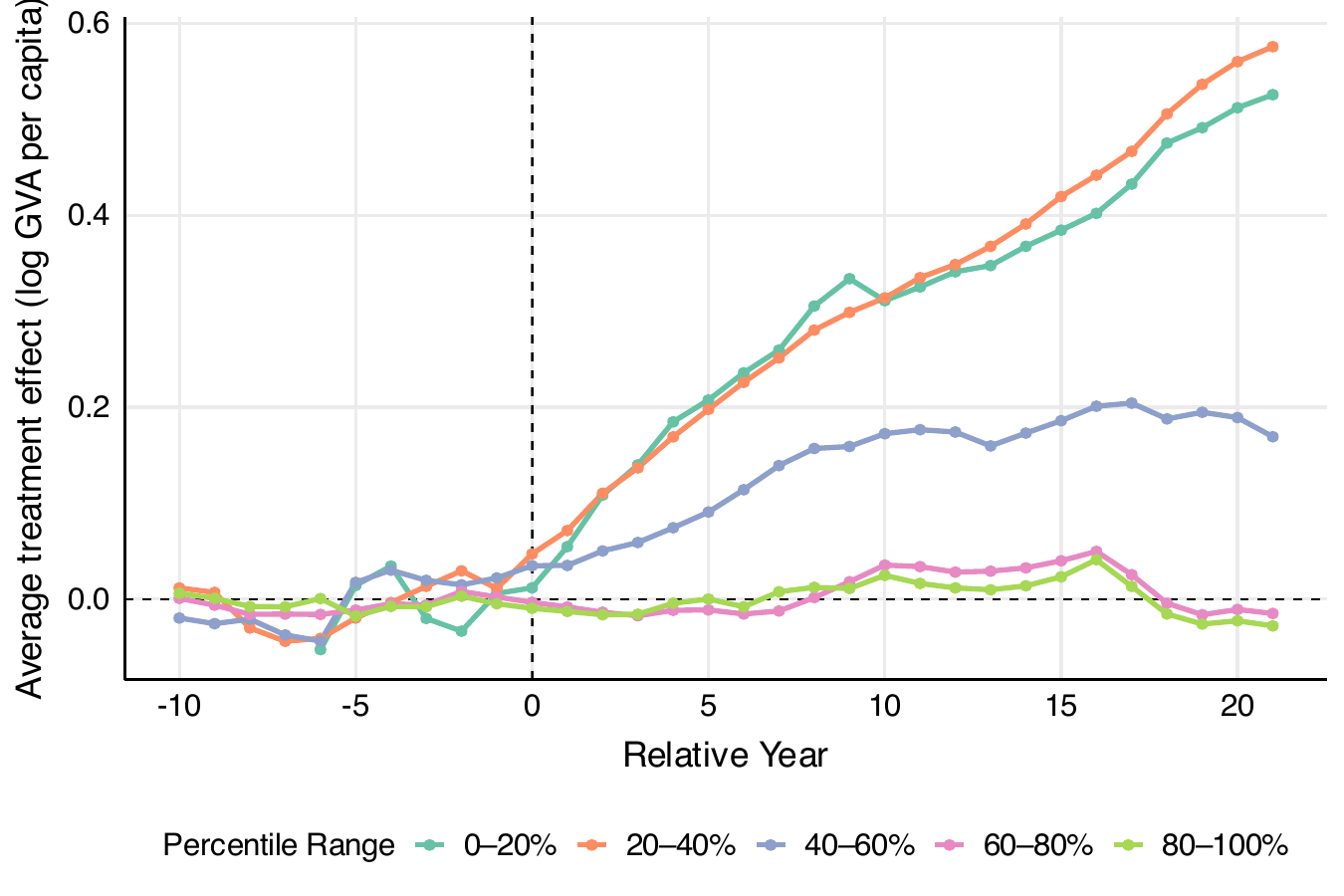}
    \vspace{-0.3cm}
    \label{fig:quant_gva}
\end{figure}

In this section we examine a different aspect of the observed heterogeneity in the estimated treatment effects. Particularly, we examine whether the effects depend on the relative position of a region across the income distribution, that is, whether poorer/richer regions experience different effects from the cohesion transfers. To do so, we categorize the treated regions into five quintiles based on their GVA per capita in the last year before the treatment (i.e., receiving CF payments for the first time). Figure \ref{fig:quant_gva} depicts $\widehat{ATT}_t$, separately for the five groups of regions. More precisely, the horizontal axis shows years relative to the start of the CF program (year 0), while the vertical axis measures the $\widehat{ATT}_t$. Each line represents a different quintile of regions ranging from the poorest $20\% (0–20\%)$ to the richest $20\% (80–100\%)$.

The graph shows substantial heterogeneity in the impact of the Cohesion Fund program across these quintiles. Regions in the lowest quintile (see red line) exhibit the largest $\widehat{ATT}_t$ in the years following treatment, while those in the highest quintile consistently display the smallest effects (see orange line). This pattern suggests that the program was significantly more beneficial for poorest regions, with treatment effects increasing over time and remaining persistently higher compared to regions in the highest quintiles. The widening gap in the estimated $\widehat{ATT}_t$ between the lowest and highest quintiles over the post-treatment period implies that the policy program contributed to the strengthening of the relative position of the poorest regions, thus ameliorating the cross-region income disparities. 

\begin{figure}[H]
    \centering
       \caption{\footnotesize{$\widehat{ATT}$ by quintile based on pre-treatment GVA per capita distribution.}}
    \includegraphics[scale=0.6]{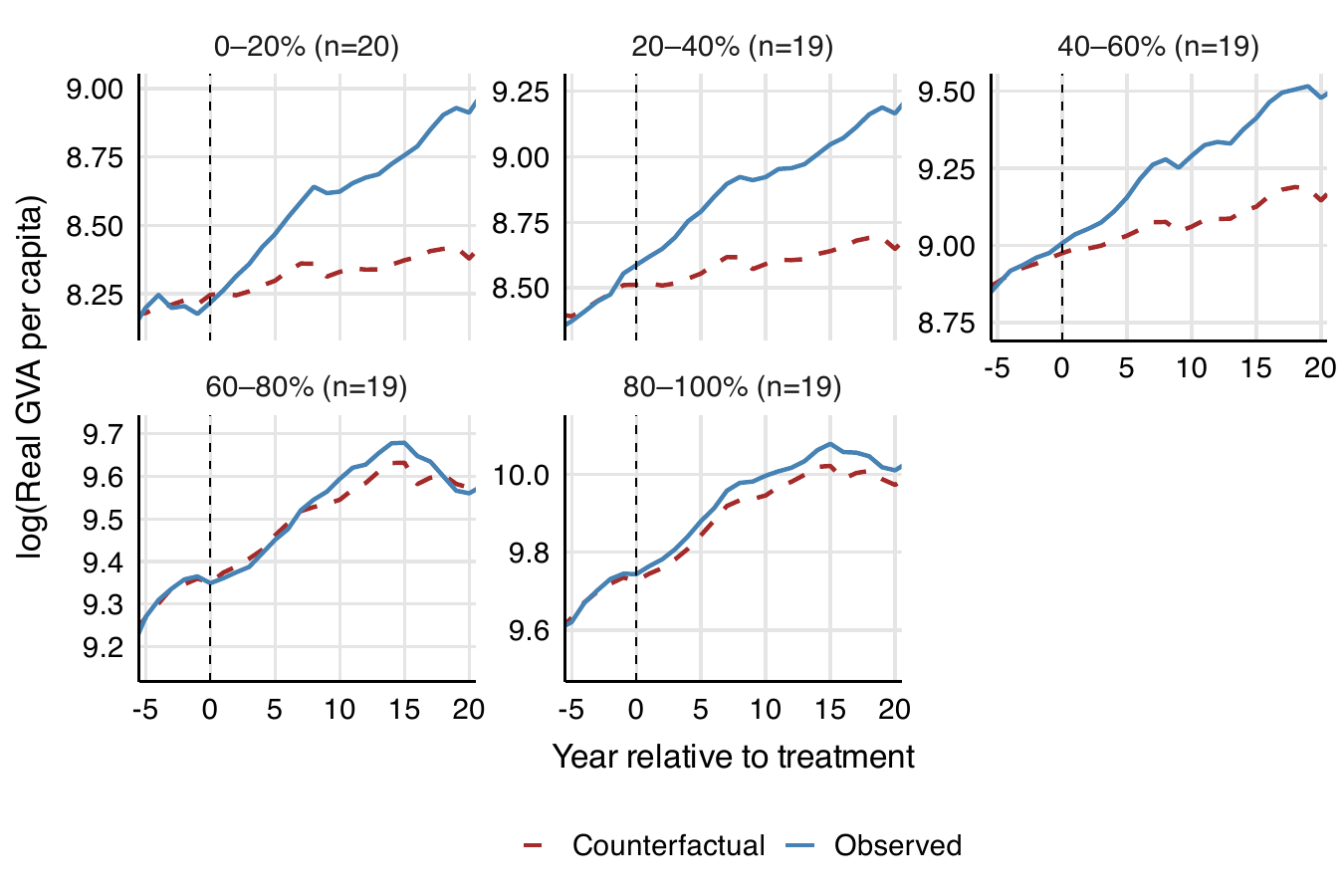}
    \vspace{-0.3cm}
    \label{fig:quant_gva_counter}
\end{figure}

Figure \ref{fig:quant_gva_counter} zooms in to examine the actual and counterfactual trajectories of log GVA per capita over time for each group of regions. The panels again group regions into quintiles based on their GVA per capita in the last pre-treatment year, from the poorest $(0–20\%)$ to the richest $(80–100\%)$. Within each group, the graph plots two series: the observed evolution of log real GVA per capita and its counterfactual (i.e., what GVA per capita would have been in the absence of the CF program). The difference between the observed (red solid) and counterfactual (light blue dashed) lines reflects the estimated treatment effect for each group. Consistent with the previous figure, the poorest regions (those in the $(0–20\%)$ and $(20–40\%)$ quintiles) display a clear and widening gap between observed and counterfactual GVA per capita in the years following treatment, indicating that they experienced substantial and sustained gains. In contrast, for richer regions (particularly those in the $(80–100\%)$ quintile), the two trajectories remain nearly parallel, suggesting minimal impact from the program. Interestingly, for the regions in the $(60–80\%)$ quintile the two trajectories coincide by the end of the period, indicating that the impact dissipates. This disaggregated analysis reinforces the earlier conclusion that the CF program had strong, persistent effects in the regions in the lower end of the income distribution. By enabling poorer regions to grow faster than their projected counterfactual paths, the program appears to have contributed significantly to reducing income disparities within the EU. Moreover, the quantitatively smaller effect estimated in richer regions suggests that the program may have limited capacity to generate impact the regions where the economic fundamentals appear relatively stronger. This could indicate a need for a more tailored design of the policy program, including a potential redistribution of funds from richer to poorer regions.

\begin{figure}[H]
    \centering
        \caption{\footnotesize{$\widehat{ATT}$ by quintile based on the distribution of the pre-treatment investment to GVA ratio.}}
    \label{fig:quant_gfcf_counter}
    \includegraphics[scale=0.6]{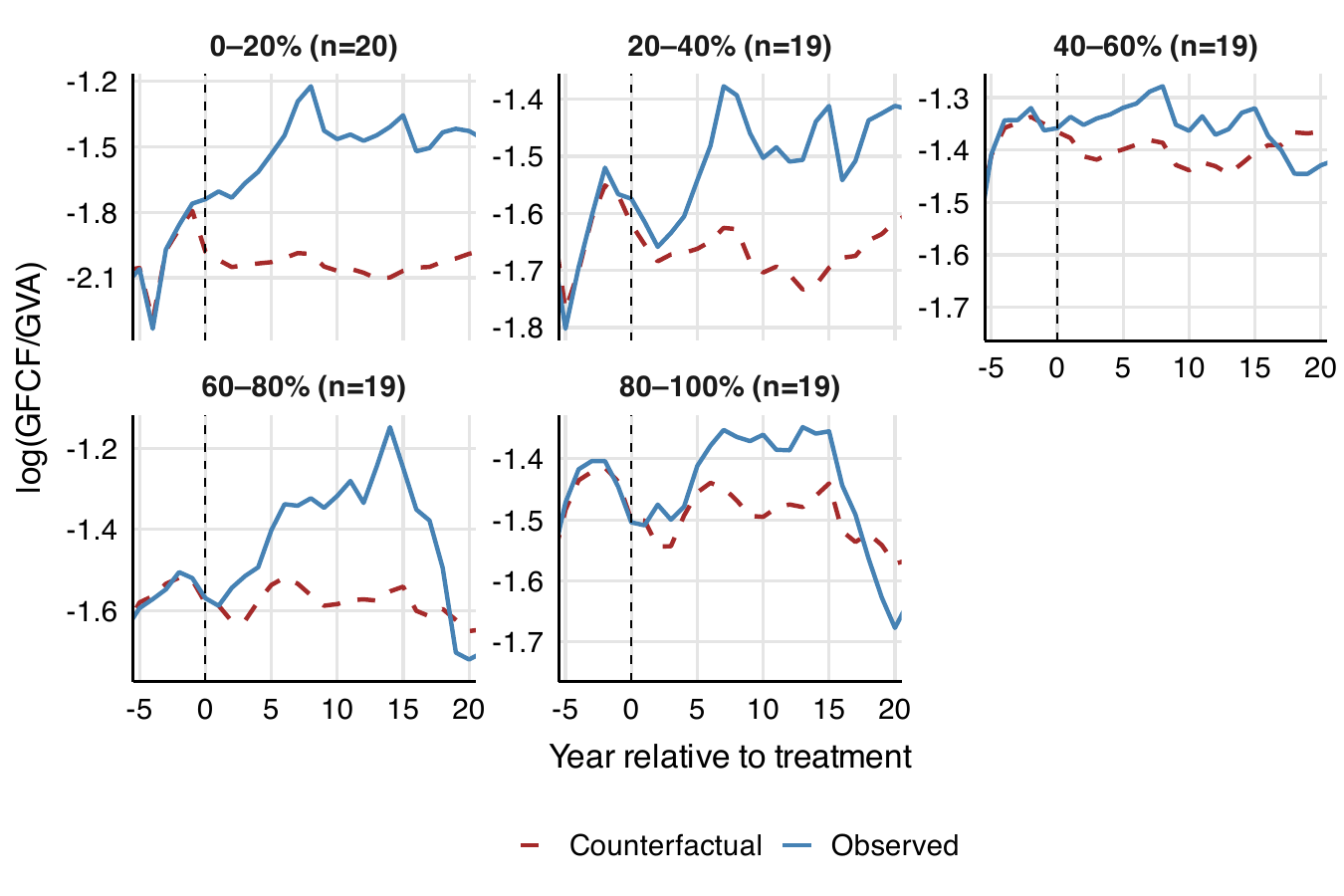}
    \vspace{-0.3cm}
\end{figure}

A similar picture emerges when we shift our focus to the evolution of investment following the disbursement of the Cohesion Fund payments. As above, Figure \ref{fig:quant_gfcf_counter} depicts the observed and the counterfactual time-path of the investment to GVA ratio in the log-scale by income quintile. As can be gleaned from the Figure, the regions belonging to the three lowest quintiles experience the largest benefits in terms of investment with the gap between observed and counterfactual widening over time (most notably in the `poorest of the poor' regions of the lowest quintile). On the contrary, in the two highest quintiles composed of the richer regions one observes that the difference between the two trajectories is rather flat, with the gap disappearing by the end of the period under consideration in $60-80\%$ percentile. This explains the negligible effect of the CF transfers in terms of GVA per capita for this group that was analyzed above.

\subsection{Regional heterogeneity over time}\label{sec:Gini}

\begin{figure}[H]
    \centering
    \caption{Observed (black solid line) and counterfactual (blue dotted line) Gini index based on GVA per capita}
    \includegraphics[scale= 0.5]{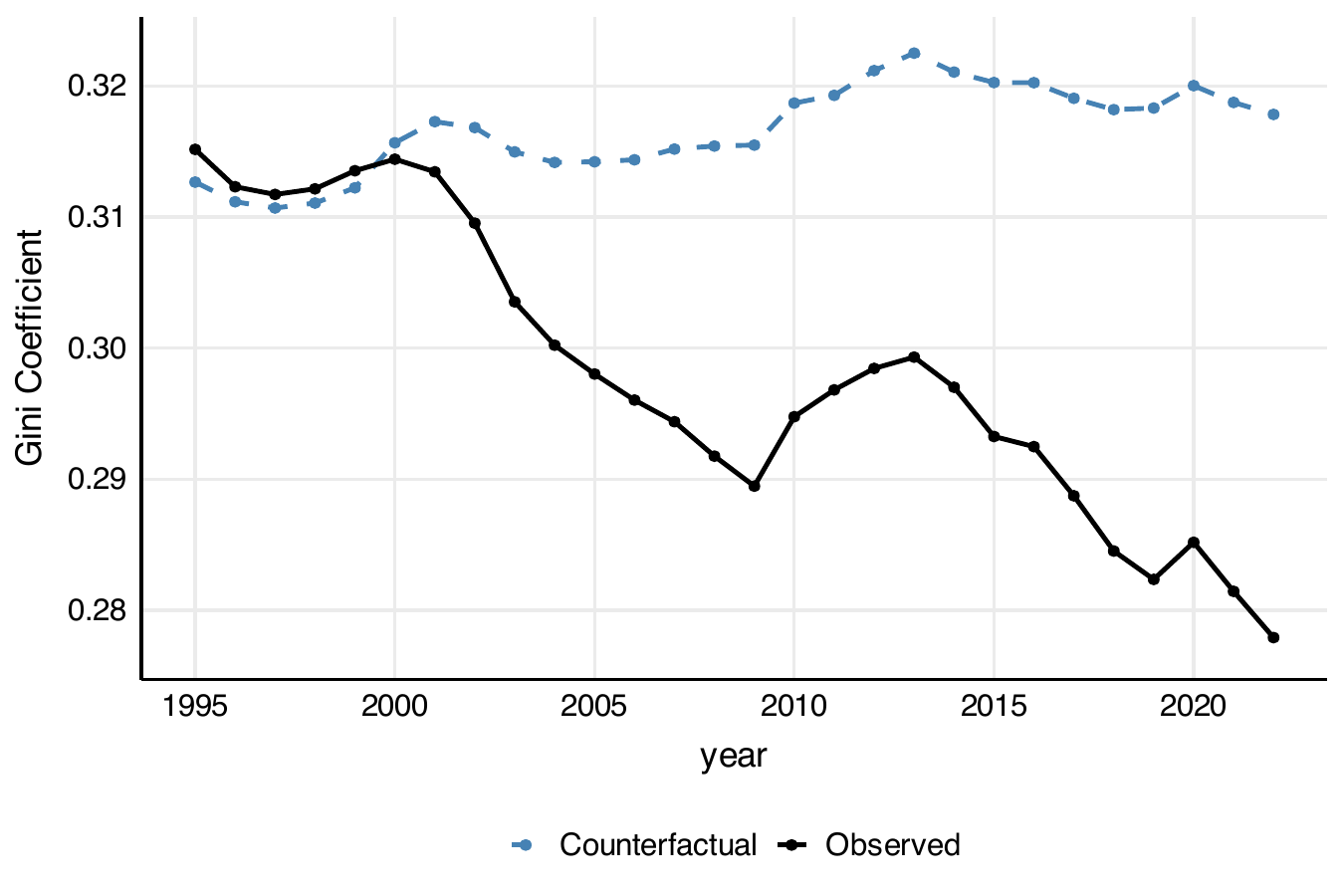}
    \label{fig:gini_gva}
\end{figure}

As shown in the previous sections, the Cohesion Fund exerts a persistent positive impact on the output of the `poorest of the poor' regions, that is, regions belonging to the bottom quintiles of the income distribution. Given these persistent effects, it is of particular interest to examine how income disparities across the EU regions have evolved over time by comparing the observed outcomes with the counterfactual scenario in which no region receives the relevant transfers. To this end, we employ one of the most commonly used statistical measures for income inequality, namely, the Gini coefficient. Specifically, we compute the time path of the Gini coefficient for both the observed (blue-dotted line) and counterfactual (red dashed-dotted line) levels of GVA per capita across all EU regions from 1995 to 2022. For the non-treated regions, the observed time series are used in both cases. We refer to this measure as the \textit{regional Gini coefficient}, as it captures the cross-regional dispersion of GVA per capita over the 1995-2022 period.  
From a visual inspection of Figure \ref{fig:gini_gva} two interesting observations emerge. First, the observed Gini coefficient declines steadily from around 0.31 to below 0.28 by the end of sample period. In sharp contrast, under the counterfactual distribution, i.e., assuming the absence of the CF program, the dispersion of GVA across regions fluctuates around 0.31 and even slightly increases over time.  

The decline begins in the early 2000s and coincides with the enlargement of the EU following the accession of the Central and Eastern European members which, as we discussed earlier, were the main recipients of the CF. The divergence between the counterfactual and actual regional Gini underlines the important role of the Cohesion Fund transfers in mitigating income disparities across the EU's regions. The strong positive effects on the output of the new Member States contributed to reducing regional divergence relative to the counterfactual no-CF scenario.

\subsection{Non-linearity}\label{sec:NLR}
\begin{figure}
    \centering
    \caption{Non-linear effect of the Cohesion Fund}
    \includegraphics[scale=0.5]{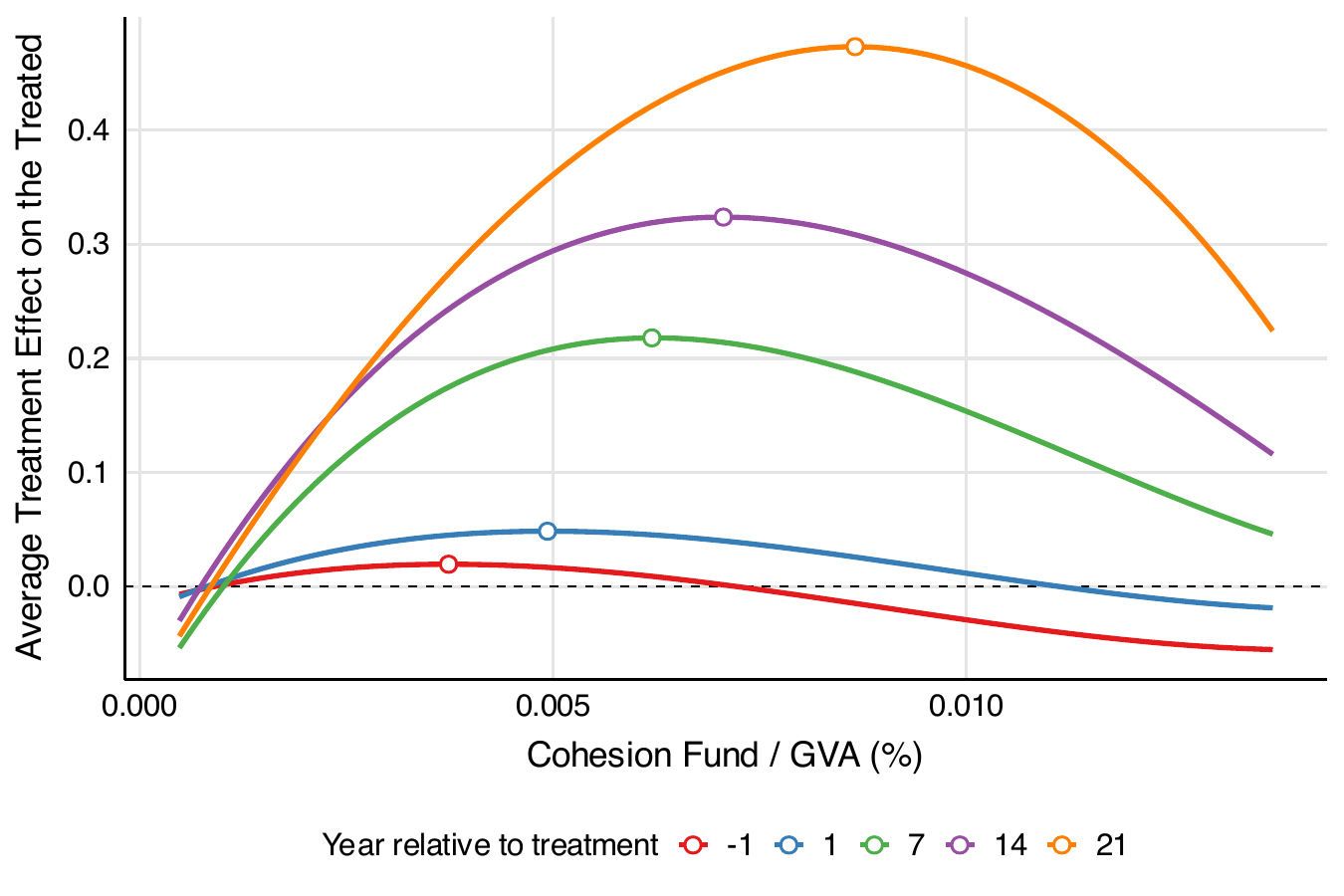}
    \label{fig:NLR_ALL1}
\end{figure}

Figure \ref{fig:NLR_ALL1} plots the estimated Average Treatment Effect on the Treated ($\widehat{ATT}$) against the intensity of Cohesion Fund transfers --measured as the share of real Cohesion Fund expenditures in real GVA-- for treated regions. Each curve corresponds to different time horizons, i.e., 1 year before treatment as well as 7, 14, and 21 years post-treatment. The relationship between treatment intensity and economic impact is clearly non-linear across all time horizons.

As expected, one year prior to the treatment, the estimated $\widehat{ATT}$ remains close to zero across all levels of funding intensity. As the post-treatment horizon extends, however, the non-linear relationship between treatment intensity and the estimated effect $\widehat{ATT}$ becomes increasingly pronounced (see the changing curvature of the lines as the post-treatment year varies from 1 to 21). Over time, the treatment intensity, i.e., the CF as a share of GVA, which generate the maximum estimated $\widehat{ATT}$-- depicted by the dots on each curve-- rises. Specifically, in year 7, the optimal treatment intensity is approximately 0.6\% of GVA, at which point the estimated effect reaches 25\%. Beyond this threshold, additional funding produces diminishing economic returns. In subsequent periods, both the optimal intensity and the peak effect rise; however, once funding exceeds the respective threshold, further transfers lead to a sharp decline in the estimated effect. The optimal treatment intensity is estimated at 0.7\% and 0.86\% for post-treatment years 14 and 21, respectively, yielding maximum $\widehat{ATT}$ effects of 0.32 and 0.47 in logs, or 35\% and 60.4\%, respectively.

These results support the hypothesis of a maximum desirable level of transfers proposed by \citet{becker2}, suggesting that while moderate levels of EU funding can stimulate growth, when transfers exceed a threshold, they can become inefficient or even counterproductive. Our results extend their analysis by emphasizing the time-varying dimension.\footnote{Our results are in line with \cite{becker2} and \cite{cerqua} who estimate that the optimal transfer intensity at 0.4\% and the maximum desirable level at 1.3\%} Specifically, Figure \ref{fig:NLR_ALL1} quantifies how the optimal level of treatment intensity --measured as funds relative to GVA-- evolves as regions progress from the initial inclusion in the program to relatively more mature stages of the program implementation. Also, the right downward-sloping segment of the curves also provides an indication of how rapidly these effects decay as funding intensity increases.\\
This analysis offers quantitative guidance for policy. That is, funding allocations should be scaled over time to remain within an effective range, avoiding the diminishing or even adverse effects observed at higher treatment intensities.

This observed shift in the treatment intensity threshold and the resulting  maximum $\widehat{ATT}$ effect is likely correlated with two factors: (i) the absorption capacity of the regions and (ii) the irregular disbursements of the funds. Overall, these results highlight the value added of the methodology for policy evaluation adopted in this paper, which allows for the calculation of time-specific effects. Previous research, see \citet{becker2}, \citet{cerqua}, provided clear evidence of a non-linear relationship for an entire programming period or combinations of periods --often constrained by limited data availability-- and therefore could not capture the temporal shift in the maximum desirable level of transfers.



\section{Conclusions}\label{sec:conclusions}

This paper studies the impact of Cohesion Fund payments on regional economic performance, focusing on the heterogeneous responses of the EU's regions to this sub-national transfers programme. In particular, relying on the matrix completion technique, it disentangles both the region- and the time-specific effects of the policy. Our main results are that these transfers exert, on average, a positive and persistent effect on both output and investment, with the majority of the gains materializing in the first seven years under treatment. Moreover, the relatively poorer regions are the ones reaping the benefits of the programme. Lastly, we uncover a non-linear relationship between funding intensity and the size of the effect.

Our results give rise to some interesting conclusions regarding the evaluation of such programmes. First and foremost, focusing on a (local) average treatment effect as the single parameter of interest masks significant information from the policymakers. Specifically, our finding that the Cohesion Fund transfers exert a larger impact on the poorer regions highlights the importance of strengthening its place-based approach. This will allow a more tailored design of the policy, focusing on the core needs of the regions. However, whether this tailored design should be the result of Member State-based designs is an issue that is left for future research.
Second, the apparent existence of a threshold beyond which the policy exhibits diminishing returns implies that there is scope for a `budget-neutral' redistribution of the Cohesion Fund payments. Specifically, as also shown in \citet{becker2}, there is scope for a redistribution of the budget that could increase the efficiency of the Cohesion Fund in terms of regional growth performance and, potentially, enhance the convergence process. 

\bibliographystyle{Chicago}
\bibliography{refs} 

\newpage

\appendix
\section*{Appendix}
\counterwithin{figure}{section}
\counterwithin{table}{section}

\section{Review of the literature}\label{app:litreview} 
Table~\ref{app:technical} provides a schematic review of the empirical literature concerned with the ex-post evaluation of cohesion policy transfers. We note that this section simply aims at providing an introduction to this voluminous literature by discussing some representative contributions; by no means should it be considered as being a thorough literature review. 

The main structural feature of this selective review will be the methodological approach employed and, secondarily, the type of data used for the cohesion transfers. More specifically, the focus in on whether a form of a policy evaluation technique (e.g. a Regression Discontinuity Design) or more traditional (panel) econometrics approach was considered.

The top panel of Table~\ref{app:technical} collects those contribution that focused on a causal inference approach. As this part was elaborated on in the main body of the paper, we briefly mention here that in terms of the empirics either an RDD or a dose response model was employed, whereas in terms of cohesion data either a binary indicator related to region eligibility or data on transfer commitments and transfer payments provided to the authors by the European Commission - Directorate General  for Regional and Urban Policy (DG REGIO).

As already mentioned in the main text, apart from the research stream focused on applying causal inference techniques for the ex-post evaluation of Cohesion policy, more traditional econometric approaches have also been utilized. 

The bulk of this literature usually adopts a panel data framework to assess the effects of cohesion policy on regional economic performance. This body of work typically exploits variation over time and across regions to estimate the long-run impacts of EU transfers. Within this strand, some studies rely on instrumental variables or spatial econometric techniques to address endogeneity concerns and account for spatial interdependencies (see, e.g., \citealt{dicaro, mohl}). Others incorporate institutional and governance dimensions into their empirical designs, highlighting how these factors condition policy effectiveness (e.g. \citealt{pose2}). 

More recent contributions shift the focus to short-run multiplier effects of cohesion spending, estimating how transfers influence regional output dynamics in the immediate aftermath of disbursement (e.g. \citealt{canova}).

\begin{landscape}

\hspace*{3.0cm} 

\noindent
\begin{minipage}{0.60\textwidth}   
\scriptsize
\setlength{\tabcolsep}{3pt}

\begin{threeparttable}

\begin{tabular}{l|lcllllc}
\toprule
\toprule
\textbf{Reference} &\textbf{Sample} &\textbf{Period} &\textbf{Data} &\textbf{Dep. Variable} &\textbf{Estimator} &\textbf{Homogeneity}  &\textbf{Spatial dimension}\\	
\midrule
\textbf{\textit{Policy evaluation for EU funding impact assessment}} &&&&&&& \\
\midrule
\textbf{\textit{RDD}} &&&&&&& \\
\citet{becker1}&674 NUTS2(p)&1989-2006&Binary&GDP pc growth&Fuzzy RDD&Homogeneous&-- \\
\citet{becker3}&186-251 NUTS2(p)&1989-2006 &Binary&GDP pc growth&Fuzzy RDD &Homogeneous (HLATE)&-- \\
\citet{becker4}&259 NUTS2&1989-2013&Binary&GDP pc growth&Fuzzy RDD &Homogeneous&-- \\
\citet{crescenzi}&NUTS2 regions in 4 countries&2000-2014&Binary&GDP pc growth&Spatial RDD& Heterogeneous (partially) &-- \\  
\citet{gagliardi}&257 NUTS2, 1233 NUTS3 &2000-2006 & Binary &GDP pc growth &Fuzzy RDD &Homogeneous  &-- \\  
\citet{pellegrini}&213 NUTS2&1994-2006&Binary&GDP pc growth&Sharp RDD&Homogeneous&--  \\
\citet{percoco}&257 NUTS2, 1233 NUTS3 &2000-2006 & Binary &GDP pc growth &Fuzzy RDD &Homogeneous(HLATE) &-- \\ 
\textbf{\textit{Dose-response}} &&&&&&& \\
\citet{becker2}&2078 NUTS3(p)&1994-2006&Payments&GDP pc growth&Dose-response (GPS)&Homogeneous&-- \\
\citet{cerqua}&208 NUTS2 &1994-2006&Payments&GDP pc growth&Dose-response (GPS)&Homogeneous&-- \\
\citet{hagen}&122 NUTS1/NUTS2&1995-2005&Payments&GDP pc growth&Dose-response (GPS)&Homogeneous&--\\
\midrule
\textbf{\textit{Panel data}} &&&&&&& \\
\midrule
\textbf{Long-run analysis} &&&&&&& \\
\cite{amendolagine} & 245 NUTS2 & 2000-2018 & Cohesion database & GDP pc growth & HSAR & Heterogeneous panel & Weighting matrix \\
\citet{bourdin} & 147 NUTS3 & 2000-2016 & Payments & GDP pc growth  & GWR, SDM & Heterogeneous panel & Weighting matrix \\
\cite{dicaro} &250 NUTS2  & 1990-2015 & Cohesion database & GDP pc & ECM & Heterogeneous panel & -- \\
\citet{fiaschi} & 175 NUTS2 & 1991-2008 &Payments, commitments & GDP pw growth & SDM & Homogeneous panel & Weighting matrix \\
\cite{fidrmuc} & 272 NUTS2 & 1997-2014 & Cohesion database & GDP pc growth & 2SLS, SDM & Homogeneous panel & Weighting matrix \\
\cite{mohl} & 126 NUTS2 & 1995-2005 & Payments & GDP pc growth & GMM, Spatial Lag & Homogeneous panel & Weighting matrix \\
\citet{pose2} & 169 NUTS2 & 1996-2007 & Payments & GDP pc growth & FE, GMM & Homogeneous panel & -- \\
\textbf{Multiplier analysis}  &&&&&&& \\  
\citet{canova} & 281 NUTS2 & 1990-2018 & Cohesion database & GVA pc growth & Local Projections & Homogeneous panel & -- \\
\citet{coelho} &238 NUTS2 &2000-2013 &Payments, Commitments &GDP pc growth &Panel IV &Homogeneous panel & \\
\citet{durand} &EU27 + UK &1994-2018 &Cohesion database &GDP pc growth & Local Projections &Homogeneous panel & \\
\citet{fiuratti} &NUTS2 of 25 EU countries &1994-2018 &Cohesion database &GDP pc growth & Local Projections &Homogeneous panel & \\
\bottomrule
\end{tabular}
			\begin{tablenotes}
                \item \textit{Note}: \textit{p} denotes that the regions are pooled for the estimation, \textit{binary} denotes a dummy variable specifying whether a region was treated, i.e. whether it receives cohesion transfers and/or is awarded the `Objective I' or `Convergence' or `less-developed' status. 
				\item \textit{Estimators}: \textit{2SLS} denotes the two-stage least squares, \textit{ECM} denotes an  error-correction model, \textit{GMM} denotes a Generalized Method of Moments estimator, \textit{GPS} denotes a generalized propensity score,  \textit{GWR} denotes a geographically weighted regression estimator, \textit{FE} denotes the fixed effects estimator, \textit{HSAR} denotes the heterogeneous spatial autoregressive estimator, \textit{SDM} denotes the Spatial Durbin Model 
				\item \textit{Data}: Cohesion data denotes the Historic EU payments database of DG REGIO mentioned in the main text, payments denotes data received by the authors directly from DG REGIO or by ESPON.
                \item \textit{Homogeneity}: Partial means that the paper includes results based on groupings of regions. HLATE denotes the case in which a heterogeneous LATE is estimate in an RDD setup, by allowing the treatment effect to vary along covaraites that do not influence the treatment status.
               \item \textit{Weighting matrix}: In order to preserve the readability of the table, the various weighting schemes employed are mentioned here: k-nearest neighbor, Gaussian distance decay function.
			\end{tablenotes}
\end{threeparttable}

\end{minipage}

\end{landscape}

\section{Technical details of empirical strategy}\label{app:technical}

\subsection*{Definitions}
Here, we provide definitions for terms used in Section \ref{sec:empirical}.
A real-valued random variable $\epsilon$ is $\sigma$-sub-Gaussian if for all $s \in \mathbb{R}$,
$$
\mathbb{E}[\exp(s\epsilon)] \leq \exp(\sigma^2 s^2 / 2).
$$
For any matrix $\mathbf{A}$ and a set of index pairs $\mathcal{O}$, define the projection matrix
$$
    (P_\mathcal{O} A)_{it} =
    \begin{cases}
        A_{it}, & \text{if } (i,t) \in \mathcal{O}, \\
        0, & \text{otherwise}.
    \end{cases}
$$
The singular value decomposition (SVD) of a $N \times T$ matrix $A$ writes $A = S \Sigma R^\top$, where $S$ and $R$ are $N \times N$ and $T \times T$ orthonormal matrices, respectively, and $\Sigma$ is an $N \times T$ rectangular diagonal matrix with singular values $\sigma_i(A)$, $i =1,\ldots,\min(N,T)$.

The Frobenius norm of a matrix $A$ is defined as
$$
\|A\|_F = \left(\sum_{i} \sigma_i(A)^2\right)^{1/2},
$$
and the nuclear norm is
$$
\|A\|_* = \sum_{i} \sigma_i(A).
$$

\subsection*{Relation of the TWFE with the interactive fixed effects models}

We assume $K=2$ latent factors in \eqref{eq:factor_model} and $\lambda_{i1} =1$, $\lambda_{i2} \in \mathbb{R}$, $f_{1t} \in \mathbb{R}$ and $f_{2t} \in \mathbb{R}$ as well as $\tau_{it} = \tau$ for each $i=1,\ldots,N$ and $t=1,\ldots,T$, then \eqref{eq:factor_model} becomes 
\begin{equation}
\label{eq:TWFE}
   y_{it} = \tau D_{it}+ f_{1t} + \lambda_{2i}  + \epsilon_{it},
\end{equation}
which is the TWFE discussed in detail, among others, by \cite{angrist2009mostly}.

\subsection*{Estimation of the low-rank matrix $L$ and $\beta$}

The minimization in \eqref{eq:min_with_covariates} is a convex problem and can be solved iteratively. A popular approach, as discussed in \cite{athey2021matrix}, is to alternate between the estimation of \( \beta \) and the low-rank matrix \( L \). That is:

\begin{enumerate}
    \item Initialize \( \beta^{(0)} \), and compute residual matrix \( R^{(0)} = Y - \mathbb{X}(\beta^{(0)}) \).
    \item Solve
    \[
        (\hat{L}^{(0)}, \hat{\Gamma}^{(0)}, \hat{\Delta}^{(0)}) = \arg\min_{L, \Gamma, \Delta} \left\{
            \frac{1}{|\mathcal{O}|} \|P_\mathcal{O} (R^{(0)} - L - \Gamma \mathbf{1}_T^\top - \mathbf{1}_N \Delta^\top)\|_F^2
            + \lambda \|L\|_*
        \right\}.
    \]
    \item Update \( \beta^{(1)} \) by solving the weighted least squares problem on the residualized data \( Y - \hat{L}^{(0)} - \hat{\Gamma}^{(0)} \mathbf{1}_T^\top - \mathbf{1}_N (\hat{\Delta}^{(0)})^\top \).
    \item Repeat until convergence.
\end{enumerate}

Each update of \( L \) can be performed via soft-thresholding of singular values using the matrix shrinkage operator as described below.

\subsection*{Matrix shrinkage and iterative soft-thresholding}

For any matrix \( A \) with SVD \( A = S \Sigma R^\top \), define the shrinkage operator
\[
    \operatorname{shrink}_\lambda(\mathbf{A}) = S \tilde{\Sigma} R^\top,
\]
where \( \tilde{\Sigma} \) replaces each singular value \( \sigma_i(A) \) with \( \max(\sigma_i(A) - \lambda, 0) \).

Using this, the soft-impute algorithm begins with
\[
    L_1 = P_\mathcal{O}(Y - \mathbb{X}(\hat{\beta}) - \hat{\Gamma} \mathbf{1}_T^\top - \mathbf{1}_N \hat{\Delta}^\top),
\]
and updates iteratively:
\[
    L_{k+1} =
    \operatorname{shrink}_{\frac{\lambda |\mathcal{O}|}{2}}
    \left\{
        P_\mathcal{O}(Y - \mathbb{X}(\hat{\beta}) - \hat{\Gamma} \mathbf{1}_T^\top - \mathbf{1}_N \hat{\Delta}^\top)
        + P_\mathcal{O}^\perp(L_k)
    \right\}.
\]

\subsection*{Tuning parameter selection}

The regularization parameter \( \lambda \) is selected via cross-validation. The set of observed entries \( \mathcal{O} \) is randomly split into \( K \) folds, and a grid of candidate values \( \lambda_1 > \lambda_2 > \cdots > \lambda_L = 0 \) is tested. The average squared prediction error on hold-out folds is computed, and the value minimizing it is selected. To improve convergence, warm-starts are used: the solution for \( \lambda_{i+1} \) is initialized using the solution from \( \lambda_i \).

\section{Data and comparisons}\label{app:data}
As already mentioned in the main text, there are two important issues related to the dataset of the Cohesion Fund expenditures.  

The first concern is related to the timing between the moment when these disbursements became financial inflows, i.e. the moment when the expenditures actually took place on the ground and the reimbursement of the funds by the European Commission to the Member States (for more on this, see \citet{ec2017}). In practical terms, this means that the dataset includes many gaps as the reimbursement of expenditures by the European Commission is not made in a timely manner. To circumvent this issue, we make use of the `modeled' expenditures series. This is a series constructed using Monte Carlo simulations that aim to approximate what the actual expenditure series may look like. 

\begin{figure}[H]
    \centering
    \caption{Correlation between modelled and raw Cohesion Fund payments}
	\includegraphics[width=0.75\textwidth]{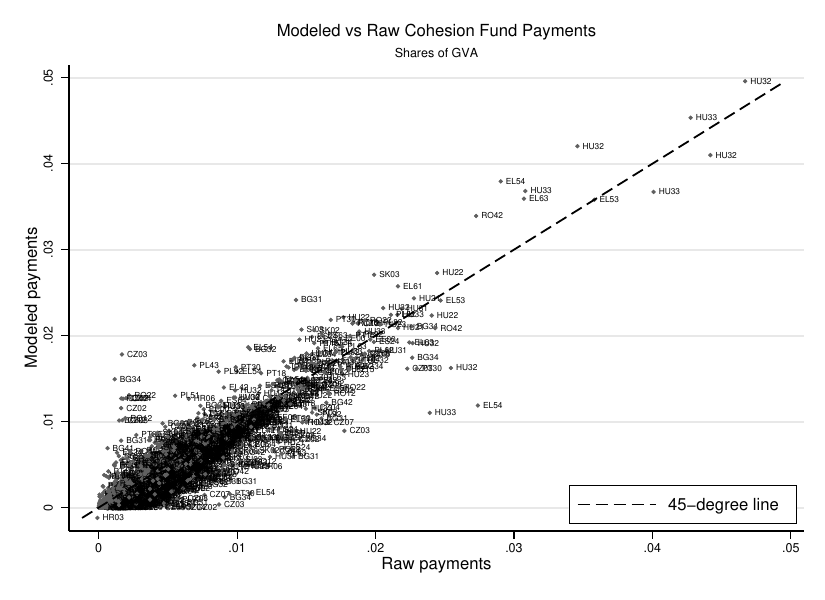}
	\caption*{\footnotesize \textit{Notes}: Scatter plot of modeled and raw Cohesion Fund expenditures. The Figure is based on a sample of 105 regions that are recipients of the Cohesion Fund and 2439 observations.}
    \label{scatter}
\end{figure}

As can be gleaned from Figure~\ref{scatter}, the correlation between the modeled and the actual payments, as shares of GVA, is --on average-- very high, with a correlation coefficient equal to 0.93 (see, also, \citet{fiuratti} for additional analysis).

Another issue is that the data are not reported under a single NUTS2 version\footnote{The Nomenclature of territorial units for statistics (NUTS), introduced in 2003, is a geographical nomenclature dividing the European Union territory into regions at three different levels. In this paper, we utilize data for the NUTS2 level given that this is the level with which EU cohesion policy is concerned. There have been 5 updates to the NUTS versions, the most recent one being introduced in 2021.}; rather, we observe that a mix of NUTS2 versions is included in the data. We proceed with converting all the data to the NUTS2 2021 version by making use of the NUTS converter developed by the European Commission's Joint Research Centre. The converter can be located in the following link: \url{https://urban.jrc.ec.europa.eu/tools/nuts-converter?lng=en}.

\section*{Comparison with TWFE}\label{COMP_TWFE}
\begin{figure}[H]
    \centering
    \includegraphics[scale = 0.5]{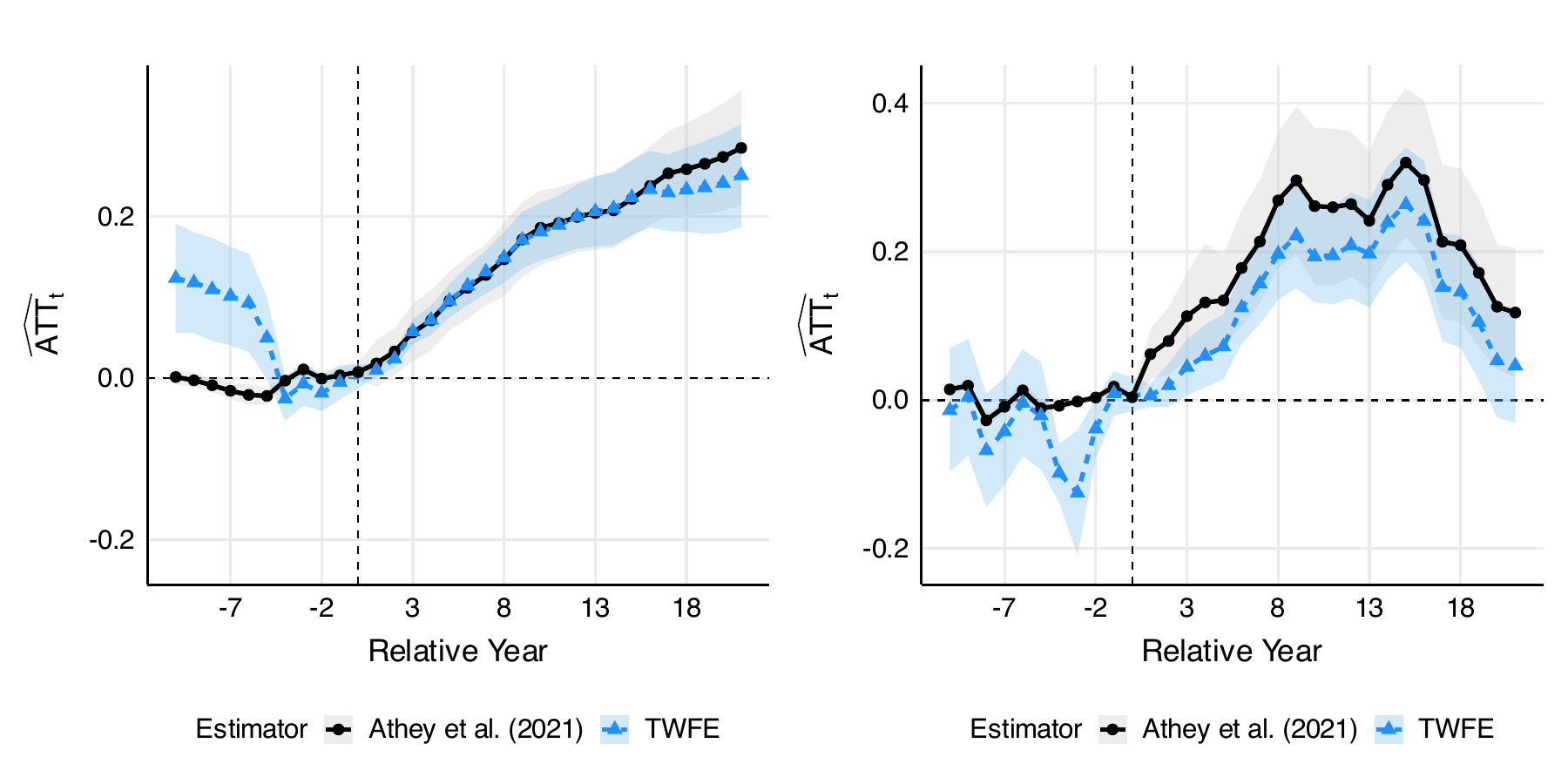}
    \caption{$\widehat{ATT}_t$ estimated by using the two way fixed effects (TWFE) against $\widehat{ATT}_t$ presented by Figures \ref{fig:ATT1} and \ref{fig:ATT2} for log-GVA per capita (left) as well as for log-gross fixed capital formation (right).  }
    \label{fig:GVA_MCNNvsTWFE}
\end{figure}

Figure \ref{fig:GVA_MCNNvsTWFE} compares the estimated Average Treatment Effects (ATT) over time using two different methodologies; the traditional Two-Way Fixed Effects (TWFE) estimator and the Matrix Completion with Nuclear Norm Minimization (MC-NNM) method proposed by \cite{athey2021matrix}. Both estimators produce qualitatively similar trajectories. The overall similarity in results lends credibility to the findings and suggests a degree of robustness in the estimated effects of the CF program.

That said, the matrix completion estimates tend to be slightly higher than those from TWFE, especially in the medium and long run. This difference is consistent with the fact that TWFE can bias estimates downward in the presence of treatment effect heterogeneity or staggered treatment adoption—both of which are relevant in this context. A key advantage of the matrix completion approach is that it does not impose constant treatment effects and instead accommodates complex patterns of response across units and over time. Importantly, the matrix completion framework also enables the heterogeneity analysis discussed in earlier figures.

\section{Tables and figures}\label{app:tables}
\subsection{Observed and counterfactual GVA levels and growth rates}

\begin{figure}[H]
    \centering
    \includegraphics[scale=0.5]{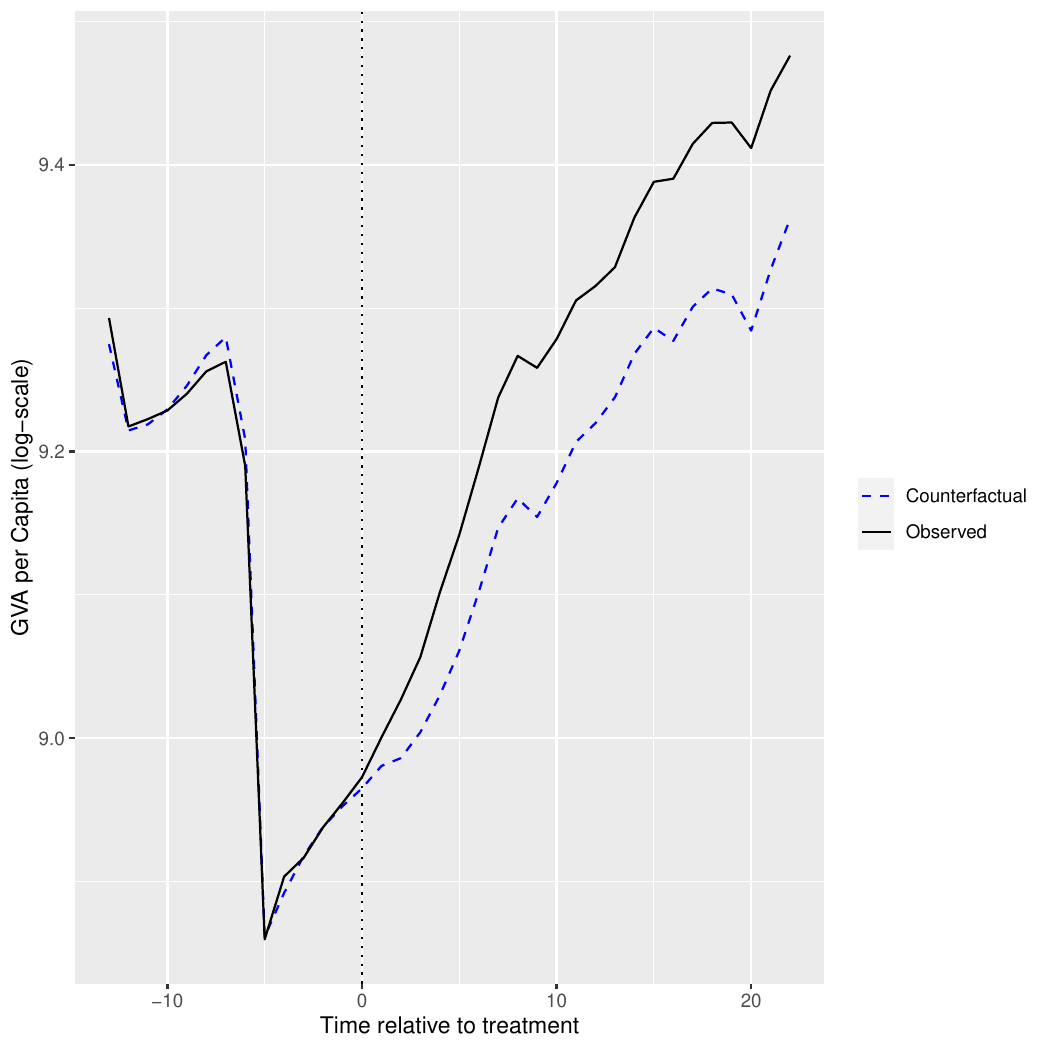}
    \vspace{-0.3cm}
    \caption{\footnotesize{Observed and estimated counterfactual average GVA in logs.}}
    \label{fig:obs_ct}
\end{figure}

\begin{figure}[H]
    \centering
    \includegraphics[scale=0.5]{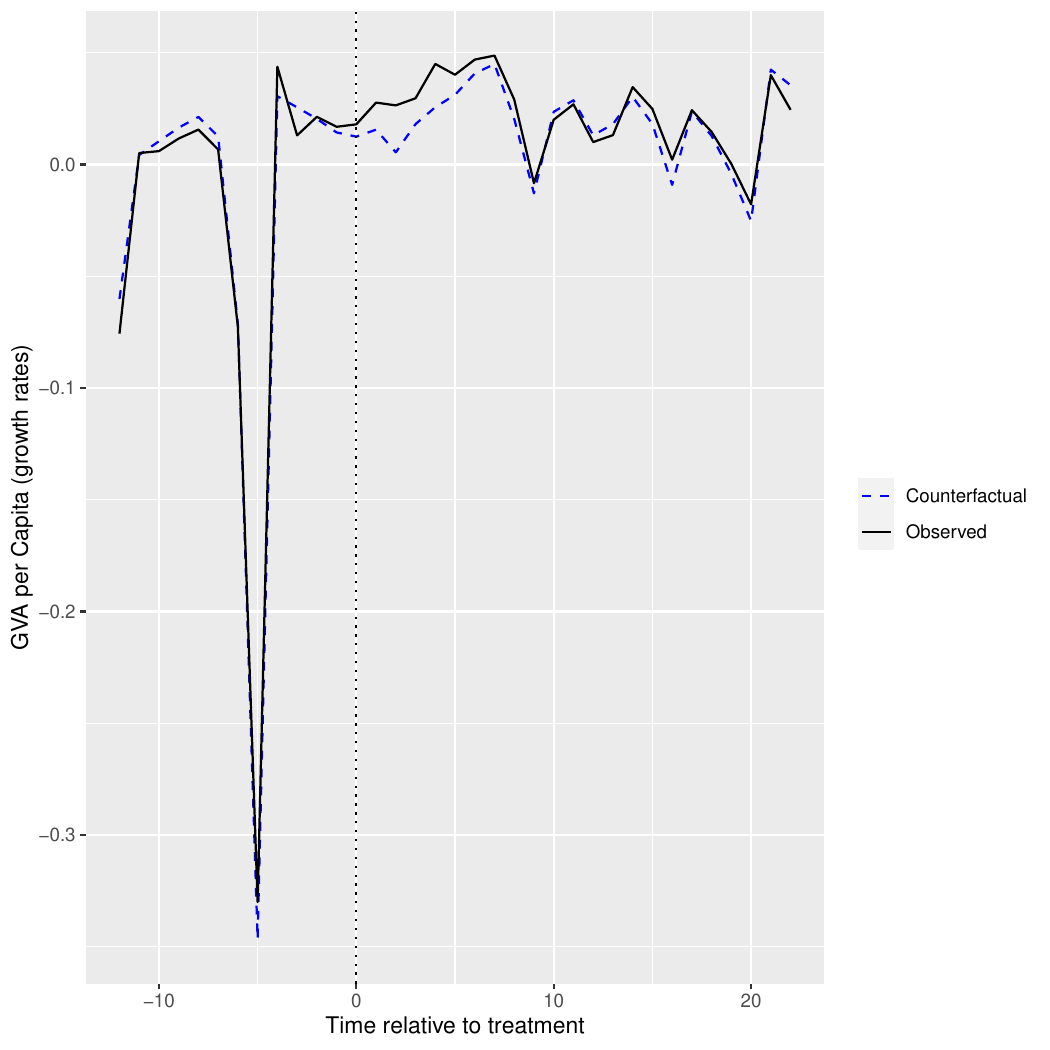}
    \vspace{-0.3cm}
    \caption{\footnotesize{Observed and estimated counterfactual average GVA growth rates.}}
    \label{fig:obs_ct_growth}
\end{figure}

\subsection{Treatment Status}

\begin{figure}[H]
    \centering
    \includegraphics[scale=0.5]{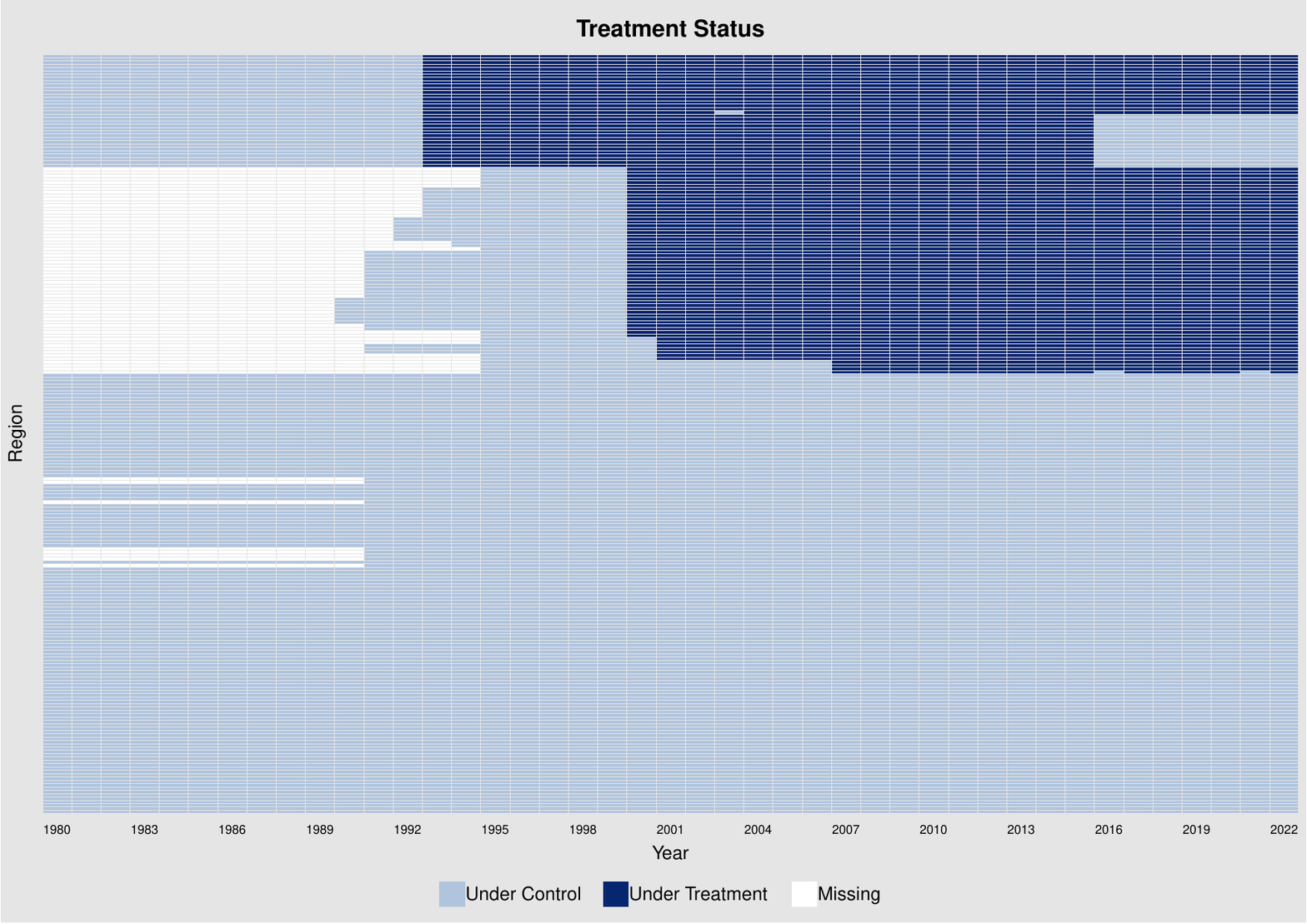}
    \vspace{-0.3cm}
    \caption{\footnotesize{Treatment Status by Region and Year}}
    \label{fig:tr_status}
\end{figure}

\clearpage

\end{document}